\documentclass[a4paper,12pt]{article}	

\usepackage{array}
\usepackage{amsmath}
\usepackage{amsfonts}
\usepackage{amssymb}
\usepackage{graphicx}
\usepackage{pifont}
\usepackage{geometry}
\usepackage{txfonts}
\usepackage{hyperref}
\usepackage{fleqn}
\usepackage{setspace}
\usepackage{xcolor}

\newcommand{\xvec}{\mbox{\bf x}}

\newcommand{\bvec}{\mbox{\bf b}}
\newcommand{\Avec}{\mbox{\bf A}}
\newcommand{\Xvec}{\mbox{\bf X}}
\newcommand{\Yvec}{\mbox{\bf Y}}

\newcommand{\zvec}{\mbox{\bf z}}
\newcommand{\yvec}{\mbox{\bf y}}

\newcommand{\disp}{\displaystyle}
\newcommand{\sigmat}{\mbox{\boldmath $\Sigma$}}
\newcommand{\muvec}{\mbox{\boldmath $\mu$}}

\begin{document}

%%%%%%%%%%%%%%%%%%%%%%%%%%%%%%%%%%%%%%%%%%%%%%%%%%%%%%%%%%%%%%%%%%%%%%%%%%%%%%%%%%%%%%%%%%%%%%%%%%%%%%%%%%%%%%%%%%%%%%%%%%%%
%%%%%%%%%%%%%%%%%%%%%%%%%%%%%%%%%%%%%%%%%%%%%%%%%%%%%%%%%%%%%%%%%%%%%%%%%%%%%%%%%%%%%%%%%%%%%%%%%%%%%%%%%%%%%%%%%%%%%%%%%%%%
\thispagestyle{empty}
\baselineskip=28pt
\vskip 5mm
\begin{center}
{\LARGE{\bf On Affine Invariant $L_p$ Depth Classifiers based\\ on an Adaptive Choice of $p$}}
\end{center}

\vskip 5mm
\baselineskip=12pt
\vskip 5mm

\begin{center}
\large
Subhajit Dutta{\footnotemark[1]} and Anil K. Ghosh{\footnotemark[2]}
\end{center}

\footnotetext[1]{
\baselineskip=10pt Department of Mathematics and Statistics, Indian Institute of Technology, Kanpur - 208016, India.\\ E-mail: duttas@iitk.ac.in} %tijahbus@gmail.com}

\footnotetext[2]{
\baselineskip=10pt Theoretical Statistics and Mathematics Unit, Indian Statistical Institute, 203, B. T. Road, Kolkata - 700108, India. E-mail: akghosh@isical.ac.in}

\baselineskip=17pt
\vskip 4mm
\centerline{\today}
\vskip 6mm

%%%%%%%%%%%%%%%%%%%%%%%%%%%%%%%%%%%%%%%%%%%%%%%%%%%%%%%%%%%%%%%%%%%%%%%%%%%%%%%%%%%%%%%%%%%%%%%%%%%%%%%%%%%%%%%%%%%%%%%%%%%%

%\begin{center}
\vspace{1.25in}
\begin{center}
{\large{\bf Abstract}}
\end{center}
In this article, we use L$_p$ depth for classification of multivariate data, where the value of $p$ is chosen adaptively using observations from the training sample. While many depth based classifiers are constructed assuming elliptic symmetry of the underlying distributions, our proposed L$_p$ depth classifiers cater to a larger class of distributions. We establish Bayes risk consistency of these proposed classifiers under appropriate regularity conditions. Several simulated and benchmark data sets are analyzed to compare their finite sample performance with some existing parametric and nonparametric classifiers including those based on other notions of data depth.

\baselineskip=16pt

\par\vfill\noindent
\par\medskip\noindent
{\bf Short title}: Affine invariant $L_p$ depth classifiers

\clearpage\pagebreak\newpage \pagenumbering{arabic}
\baselineskip=24pt

%%%%%%%%%%%%%%%%%%%%%%%%%%%%%%%%%%%%%%%%%%%%%%%%%%%%%%%%%%%%%%%%%%%%%%%%%%%%%%%%%%%%%%%%%%%%%%%%%%%%%%%%%%%%%%%%%%%%%%%%%%%%

\section{Introduction}
%\label{Intro}

Data depth measures the centrality of a point $\xvec$ in $\mathbb{R}^d$ with respect to a $d$-dimensional data cloud, or a $d$-dimensional probability distribution. As a result, it provides a centre-outward ordering of multivariate data. Various notions of data depth are available in the literature (see, e.g., Liu, Parelius and Singh, 1999; Zuo and Serfling, 2000), and they have been used for generalizing many univariate statistical methods to the multivariate setup. One important application is supervised classification.
Ghosh and Chaudhuri (2005) introduced maximum depth classifiers and also developed a modified classifier based on Tukey's (1975) half-space depth (HD). Later, Dutta and Ghosh (2012) investigated some robust classifiers based on projection depth (PD) (see, e.g., Zuo and Serfling, 2000) and robust versions of Mahalanobis depth (MD). Li, Cuesta-Albertos and Liu (2012) developed nonparametric classifiers based on depth-depth (DD) plots. Other depth based classification methods include the work of J\"{o}rnsten (2004), Hoberg and Mosler (2006), Hartikainen and Oja (2006), Cui, Lin and Yang (2008) and Paindaveine and Van Bever (2015).

Constructions of most of these classifiers were motivated by elliptic symmetry (i.e., $l_2$-symmetry after an affine transformation; see, e.g., Fang, Kotz and Ng, 1989) of the competing class distributions. In this article, we develop some classifiers motivated by general $l_p$-symmetry (after an affine transformation) of the underlying distributions. We assume that the density of each competing class to be continuous, and of the form $f(\xvec)= \psi(\|{\bf A}(\xvec -{\bf b})\|_p)$, where $p \ge 1$, ${\bf b}$ is a $d$-dimensional vector, ${\bf A}$ is a $d \times d$ non-singular matrix and $\psi : \mathbb{R}_+ \rightarrow \mathbb{R}_+$ is a scalar continuous function. Here, $\|\zvec\|_p=(|z_1|^p+\cdots+|z_d|^p)^{1/p}$ for any $\zvec=(z_1,\ldots,z_d)^T \in \mathbb{R}^d$. Clearly, the location (i.e., the centre of symmetry) $\muvec$ of this distribution is ${\bf b}$. In the case of $p=2$ (i.e., elliptic symmetry), the associated scatter matrix $\sigmat$ is given by $({\bf AA}^{T})^{-1}$. If $f$ is assumed to have finite second moments, it has the mean vector $E[\Xvec]={\bf b}$ and the dispersion matrix $E[(\Xvec-\muvec)(\Xvec-\muvec)^{T}]=c_{p, \psi}~{({\bf AA}^{T})}^{-1}$, where $c_{p, \psi}$ is a positive constant that depends on $p$ and $\psi$. So, ${\bf A}$ can be viewed as a square root of $\sigmat^{-1}$ (upto a scalar constant). Several authors have studied various properties of multivariate $l_p$-symmetric distributions (see, e.g., Yue and Ma, 1995; Gupta and Song, 1997). Sinz and Bethge (2010) used $l_p$-symmetric distributions to develop various statistical tools including independent component analysis.  Arellano-Valle and Richter (2012) have introduced skewed versions of $l_{p}$-symmetric distributions. Throughout this article, by $l_p$-symmetric, we mean $l_p$-symmetric after applying an affine transformation (as we have discussed above).
%Throughout this article, we define $\sigmat^{-1/2}$ as the symmetric square root of $\sigmat$. If $\sigmat$ is positive definite, this choice of $\sigmat^{-1/2}$ is unique (see, e.g., Theorem 7.2.6 in Horn and Johnson, 1994).

If the underlying class distributions are $l_2$-symmetric (i.e., spherically or elliptically symmetric), for several existing notions of data depth, the class densities turn out to be functions of depths.
% So, the depth contours coincide with the corresponding density contours, and
Consequently, the Bayes classifier can be expressed as a function of data depths corresponding to different competing classes. This is the main argument used in proving the Bayes risk consistency for most of the existing depth based classifiers. However, if the depth contours fail to match the density contours, one cannot use such mathematical arguments. To appreciate this, consider a two-class problem with equal priors and class densities $f_1(\xvec) = c_p \exp(-\| \xvec \|_p^p)$ and $f_2(\xvec) = c_p/\sigma^d \exp(-\| \xvec \|_p^p/\sigma)$, where $\sigma \neq 1$ is a positive constant, and $c_p$ is the normalizing constant. Clearly, the boundary of the Bayes classifier is of the form $\{\xvec : \|\xvec\|_p = K_0\}$ for some $K_0>0$. Now, MD contours for each class are concentric hyper-spheres of the form $\{\xvec : \|\xvec\|_2 = K\}$ for varying choices of $K > 0$. Therefore, when $p \neq 2$, no classifier based on MD can yield the Bayes class boundary.

\begin{figure}[h]
\vspace{.1 in}
\begin{center}
\begin{tabular}{c c c}
% \hspace{-0.2in}
% \vspace{-.1 in}
\includegraphics[width=4.70cm]{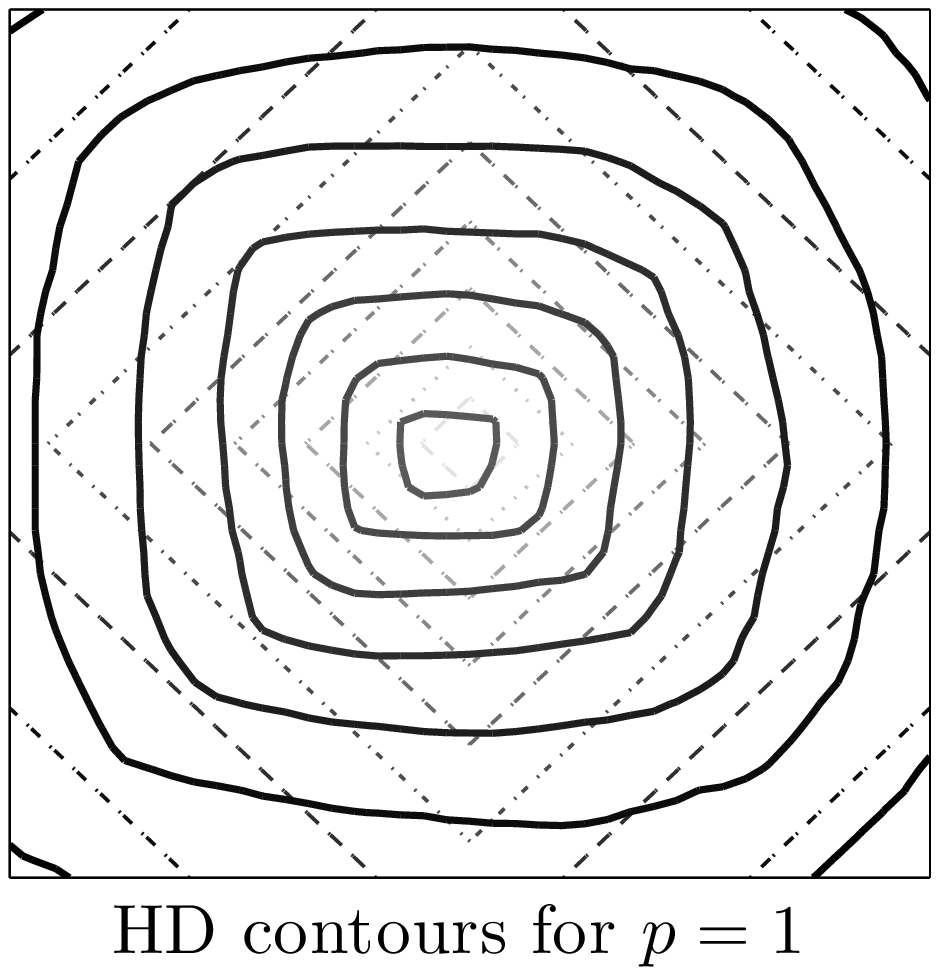}&
\includegraphics[width=4.70cm]{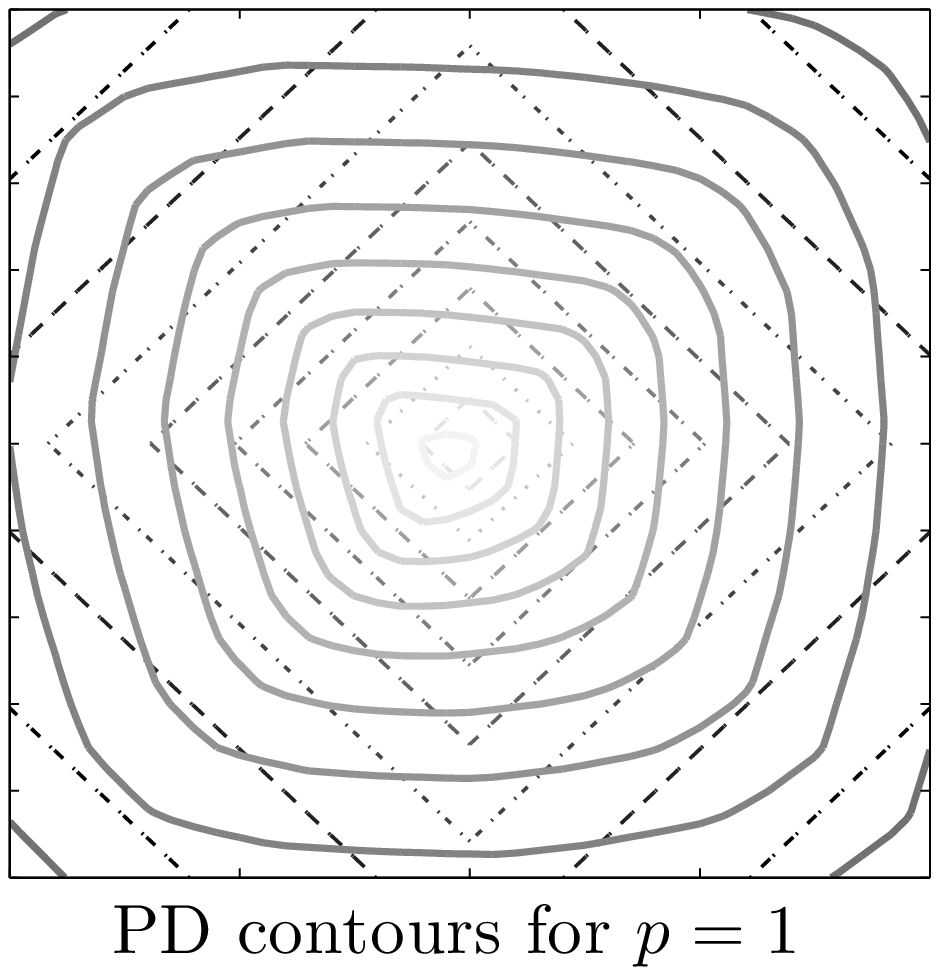}&
\includegraphics[width=4.70cm]{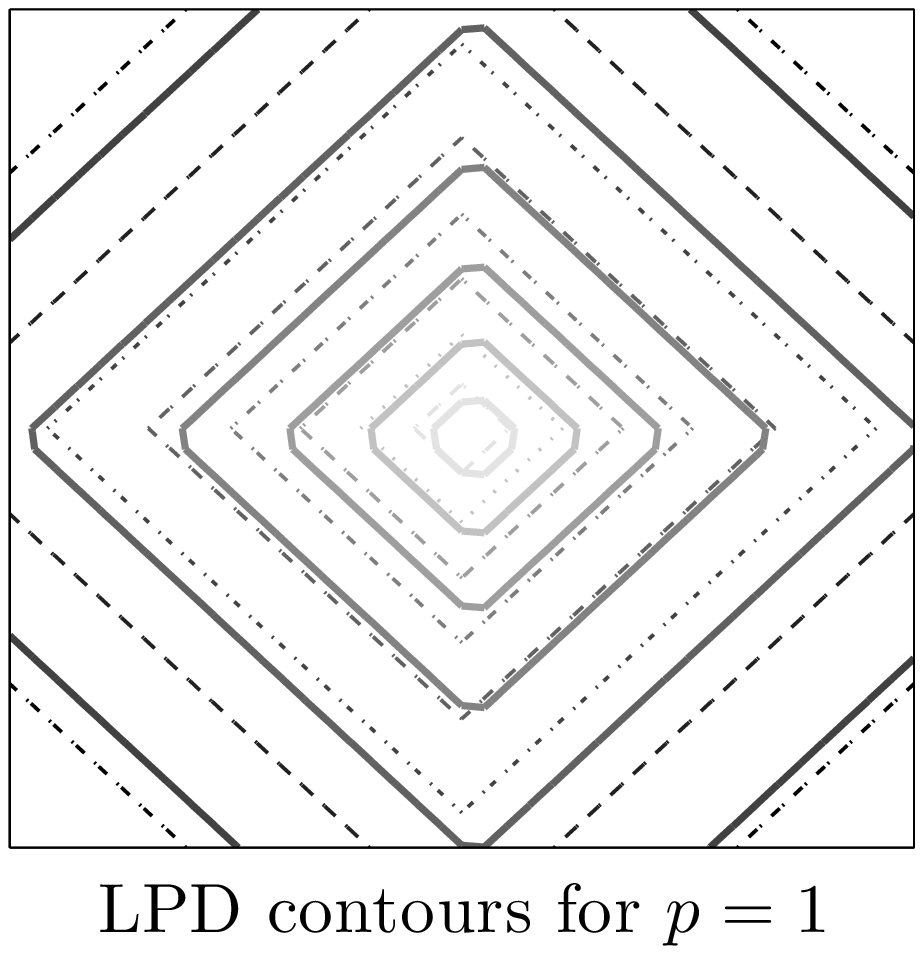}\\
\includegraphics[width=4.70cm]{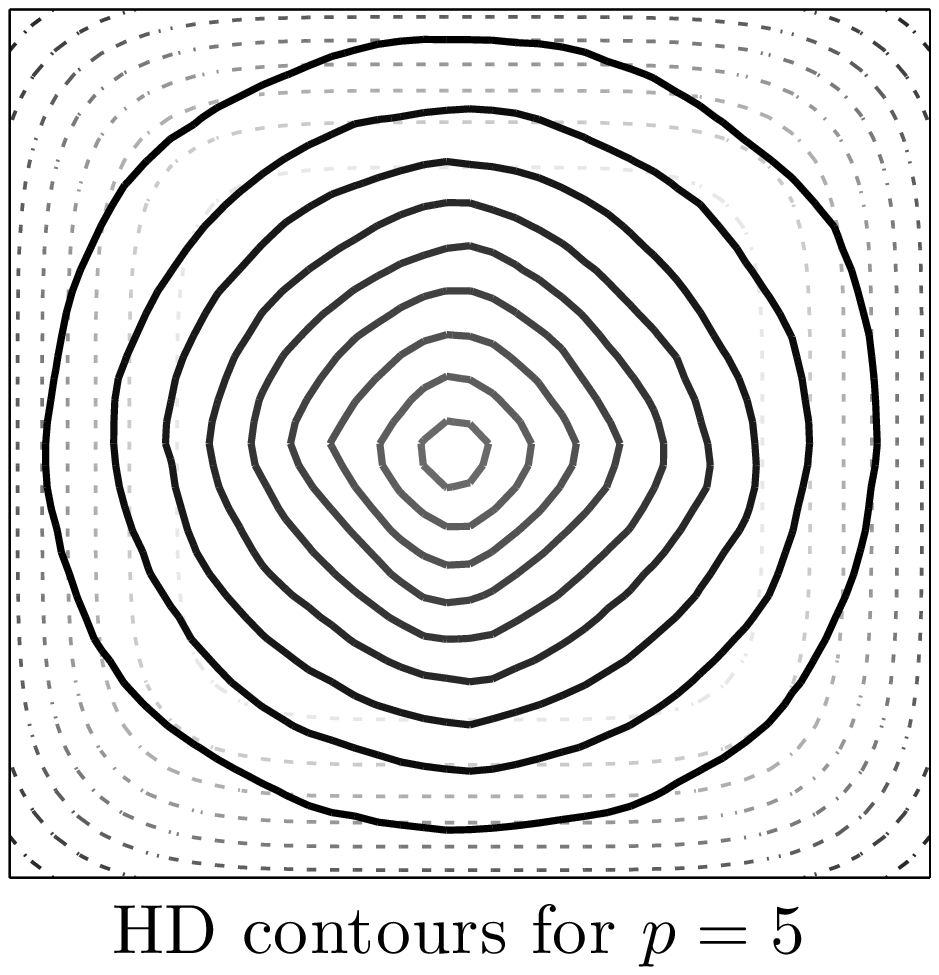}&
\includegraphics[width=4.70cm]{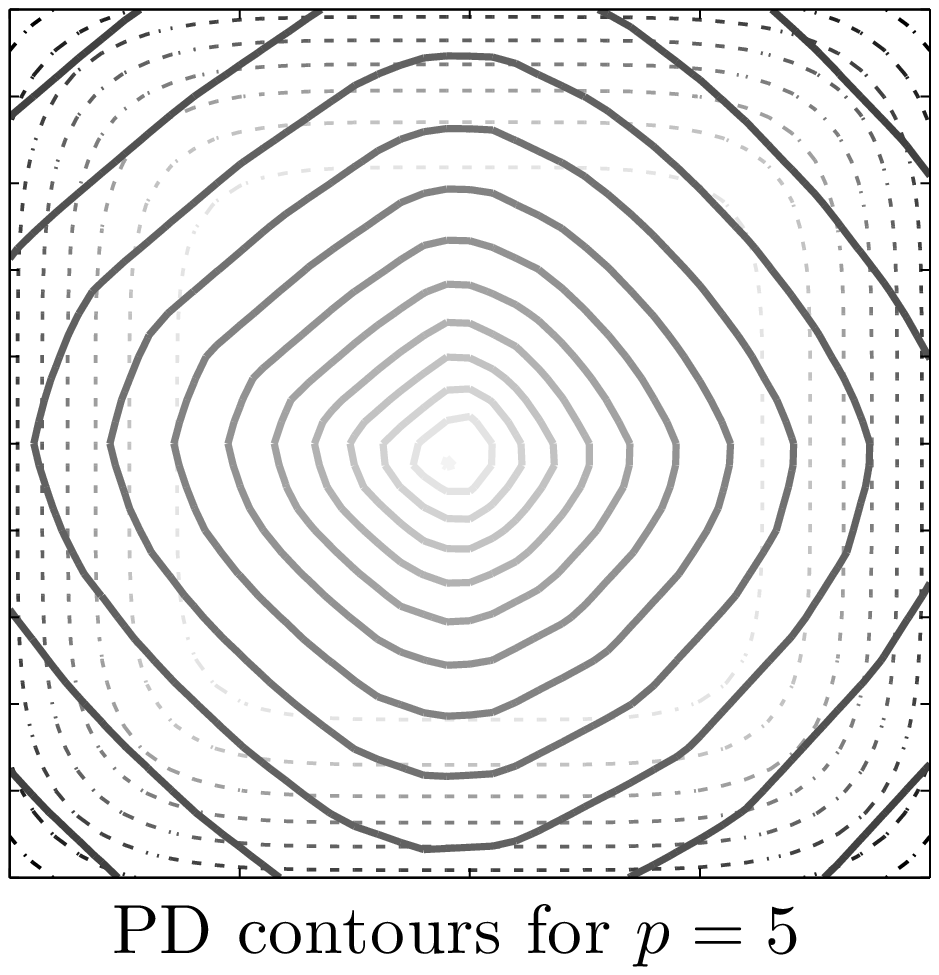}&
\includegraphics[width=4.70cm]{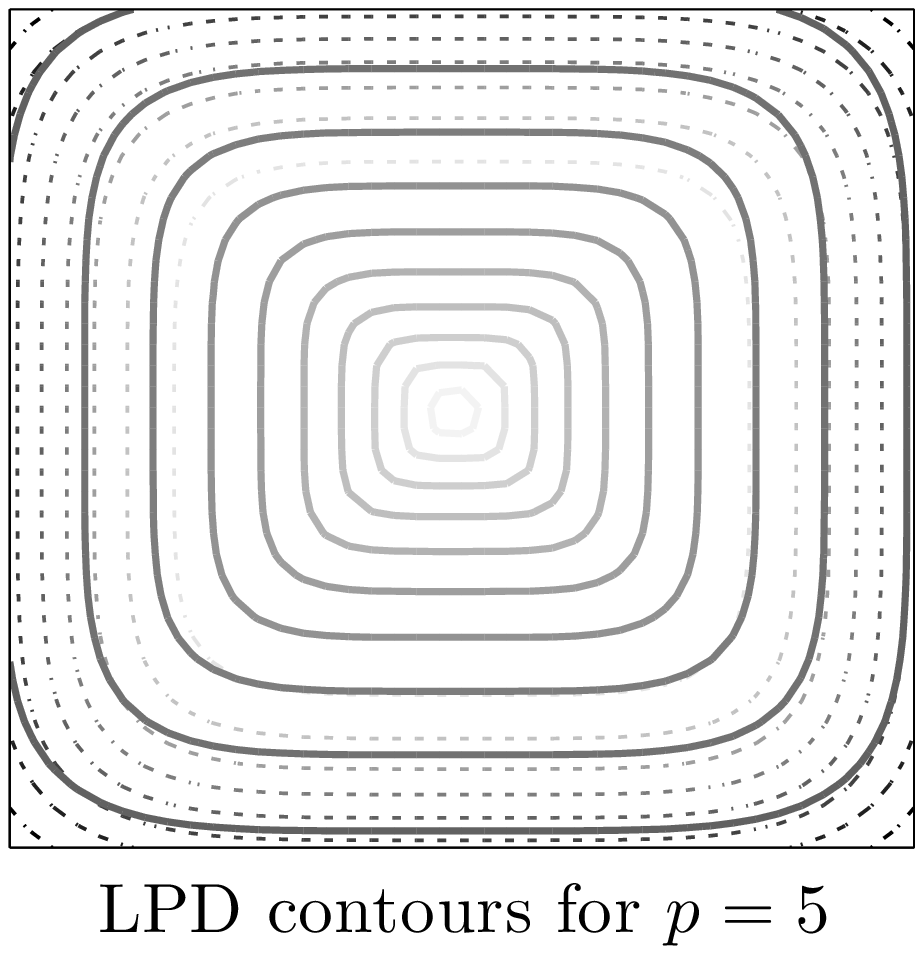}\\
\end{tabular}
 \vspace{-.05 in}
\caption{Density contours ({\tt dotted curves}) of $l_p$ symmetric distributions for $p=1$ and $p=5$, and the corresponding HD, PD and L$_p$D contours ({\tt bold curves}) estimated from the data.}
\end{center}
 \vspace{-0.15in}
\end{figure}

\noindent
For $l_p$-symmetric distributions with $p \neq 2$, Dutta, Ghosh and Chaudhuri (2011) proved that the density cannot be a function of HD as well. Figure 1 shows HD and PD contours (indicated using {\tt bold curves}) computed based on 2000 observations from two $l_p$-symmetric densities (density contours are shown using {\tt dotted curves}) with $p=1$ and $p=5$ (see the left and the middle panels). This figure clearly shows that the class density cannot be a function of depth in either of these cases.
% Therefore, in such cases, the classifiers based on empirical versions of PD, MD and HD may fail to yield satisfactory results.

To overcome this limitation, we use L$_p$ depth (see, e.g., Zuo and Serfling, 2000) with a data driven choice of $p$. L$_p$ depth (L$_p$D) of an observation $\xvec$ with respect to a multivariate distribution function $F$ is defined as $\delta_p(\xvec)=[1+r_p(\xvec)]^{-1}$, where $r_p(\xvec)=\|\sigmat^{-1/2}(\xvec-\muvec)\|_p$ with $\muvec$ and $\sigmat$ being the location and the scatter associated with $F$. The empirical version of $\delta_p(\xvec)$ is obtained by estimating $p$, and $r_p(\xvec)$ from the data (see Section 2 for details). The right panel of Figure 1 shows that the empirical L$_p$D contours based on 2000 observations almost coincide with the underlying $l_p$-symmetric density contours both for $p=1$ and $p=5$. So, classifiers based on $\delta_p(\xvec)$ are expected to yield an improved performance.

\section{Estimation of L$_p$ Depth}

To construct a classifier based on $\delta_p(\xvec)$, first one needs to find an appropriate value of $p$ for each of the competing classes. We choose the value of $p$ which fits to the data well. Suppose $\xvec_{1}, \xvec_{2}, \ldots, \xvec_{n}$ are $n$ independent observations from a $l_p$-symmetric distribution with density $f$. If ${\bf b}$ and ${\bf A}$ denote the associated location and scatter parameters (defined in Section 1), then $f$ can be expressed as:
\begin{equation*}
f(\xvec, p) = |{\bf A}| ~\frac{p^{d-1}\Gamma (d/p)}{2^d [\Gamma (1/p)]^d} \frac{g_p(\delta_{p}(\xvec)) ~\delta_p(\xvec)^{d+1}}{{[1-\delta_{p}(\xvec)]}^{d-1}}, \nonumber
%\footnote{\color{red} modify expression for $g_p$ \color{blue} done}
%\vspace{-0.0in}
\end{equation*}
where $\delta_p(\xvec)=[1+\|{\bf A}(\xvec-{\bf b})\|_p]^{-1}$, and  $g_p$ is the density of $\delta_p(\Xvec)$ (see Lemma 1 in the Appendix). %Throughout this article, we assume that $g_p(\cdot)$ is uniformly continuous.
In Section 2.1, we discuss the estimation procedure for $p$, assuming that ${\hat {\bf b}}$ and ${\hat {\bf A}}$ (estimates for ${\bf b}$ and ${\bf A}$, respectively) are given. In Section 2.2, we construct ${\hat {\bf b}}$ and ${\hat {\bf A}}$ which makes the estimate of $p$, and hence the empirical version of L$_p$D affine invariant.

\subsection{Estimation of $p$}

For any fixed $p$, using the data $\xvec_1,\xvec_2,\ldots,\xvec_n$ from $f$, we first compute $\hat {r_{p}}(\xvec_{i})=\|{\hat {\bf A}}(\xvec_i-{\hat {\bf b}})\|_p$ and ${\hat \delta}_p(\xvec_i)=[1+\hat {r_{p}}(\xvec_{i})]^{-1}$ for $i=1,2,\ldots,n$. %, where ${\hat {\bf A}}$ and ${\hat {\bf b}}$ are the estimates of ${\bf A}$ and ${\bf b}$ computed from the data. %Note that if we assume $f$ to be $l_p$-symmetric,
In order to estimate $f$, one needs an estimate of $g_p$.
Assuming $\hat {\delta_{p}}(\xvec_{1}), \ldots, \hat {\delta_{p}}(\xvec_{n})$ as sample observations, we estimate it using the kernel density estimation (see, e.g., Silverman, 1998) method. This density estimate is given by $\hat g_{p,h}(\delta) = \frac{1}{nh} \sum_{i=1}^{n} K \left [ h^{-1}({\delta - \hat {\delta_{p}}(\xvec_{i})}) \right ]$, where $K$ is the kernel function and $h$ is the associated bandwidth parameter. %Throughout this article,
We use the Gaussian kernel $K(t)=(2\pi)^{-1/2}e^{-t^2/2}$, and $h$ is chosen using the bandwidth selection method proposed in Sheather and Jones (1991).
So, under the assumption of $l_p$ symmetry, the estimate of the density function $f$ is given by
\begin{equation*}
\hat f_h(\xvec, p)=|\hat {{\bf A}|} ~\frac{p^{d-1}\Gamma (d/p)}{2^d \{\Gamma (1/p)\}^d} \frac{\hat g_{p,h}(\hat {\delta_{p}}(\xvec))~
\hat {\delta_{p}}(\xvec)^{d+1}}{ [1-\hat \delta_{p}(\xvec)]^{d-1}}. \nonumber
\end{equation*}
Irrespective of the dimension of the data, here we need only one-dimensional kernel density estimation. %This helps us to get rid of the curse of dimensionality that one usually faces in high-dimensional nonparametric density estimation.
Similar approaches for depth based density estimation were also used by Fraiman, Liu and Mechole (1997) and Hartikainen and Oja (2006).

% To find an appropriate value of $p$,
We compute the estimated joint likelihood ${\cal L}_{p}(\xvec_1, \ldots, \xvec_n)$ $=\prod_{i=1}^{n} \hat f_{h}(\xvec_{i},$ $ p)$ for different $p$, and choose the one that maximizes ${\cal L}_{p}(\xvec_1, \ldots, \xvec_n)$ or $\log {\cal L}_{p}(\xvec_1, \ldots, $ $\xvec_n)$. However, note that if $\hat \delta_p(\xvec)$ is close to zero or unity, $|\log \hat f_h(\xvec, p)|$  tends to be very influential. So, we consider only those $\xvec_i$s for which $\hat \delta_p(\xvec_i)$ lie between $\zeta_{1n}$ and $\zeta_{2n}$ ($0<\zeta_{1n}<\zeta_{2n}<1$) and find $\hat p$ (the estimate of $p$) by maximizing
$${\varphi}_{p}(\xvec_1, \cdots, \xvec_n)=\sum_{\{i~:~\hat \delta_p(\xvec_i) \in [\zeta_{1n},\zeta_{2n}]\}} \log \hat f_{h}(\xvec_{i}, p)$$
over ${\cal P}=\{p_1,p_2,\ldots,p_M\}$, a finite set of values for $p$. The following theorem suggests a suitable choice for $[\zeta_{1n},\zeta_{2n}]$, and gives the asymptotic behavior of $\hat p$ under appropriate regularity conditions.

\vspace{0.05in}
{\bf Theorem 1:} {\it Let $\Xvec_1,\Xvec_2,\ldots,\Xvec_n$ be independent and identically distributed with the density of the form $f(\xvec, p_0)=\psi_0(\|{\bf A}(\xvec-{\bf b})\|_{p_0})$ for some
% ${\bf b} \in\mathbb{R}^d$, a $d \times d$ nonsingular matrix ${\bf A}$,  $\psi_0:\mathbb{R_+}\rightarrow \mathbb{R_+}$ and
$p_0 \geq 1$. For any $p \geq 1$, define $\zeta_{1n}=\zeta_1(p)$ and $\zeta_{2n}=\zeta_2(p)$ as the $\alpha_{1n}$-th and $\alpha_{2n}$-th quantile of $g_p$, where $\alpha_{1n}\downarrow0$, $\alpha_{2n}\uparrow1$ and $n^{1/2}~\min\{\zeta_{1n}, 1-\zeta_{2n}\} \rightarrow \infty$ as $n \rightarrow \infty$. Now, consider the following assumptions:

\noindent
(C1) ${\hat {\bf A}}$ and ${\hat {\bf b}}$ are $\sqrt{n}$-consistent estimates of $a_{0}{\bf A}$ and ${\bf b}$ (i.e., $\sqrt{n}\|{\hat {\bf A}}-a_0A\|_F=O_P(1)$, where $a_{0} > 0$ and $\|\cdot\|_F$ denotes the Frobenious norm; and $\sqrt{n}\|{\hat {\bf b}}-{\bf b}\|_p=O_P(1)$ for any $p\ge 1)$, respectively.

\noindent
(C2) For any $p\ge 1$, $g_p$ is absolutely continuous and the bandwidth $h$ associated with the estimation of $g_p$ is of the order  $O(n^{-1/4+\varepsilon})$ for some $\varepsilon \in (0, 1/4)$.

\noindent
(C3) For any $p\ge 1$, $\disp n^{1/2}h^2 \inf_{\{\delta_{p}(\xvec) \in~ [\zeta_{1n},\zeta_{2n}]\}} g_p(\delta_{p}(\xvec)) \rightarrow \infty$ as $n \rightarrow \infty$.

\noindent
Assume (C1)-(C3), and define $\hat p_n= \arg\max_{p} {\varphi}_{p}(\Xvec_1, \cdots, \Xvec_n)$. If $p_0 \in {\cal P}$, then $\hat p_n \stackrel{P}{\rightarrow} p_0$ as $n \rightarrow \infty$. If $p_0 \notin {\cal P}$, then $\hat p_n \stackrel{P}{\rightarrow}p_0^*$, where $p_0^*$ minimizes the Kullback-Leibler divergence between $f(\cdot, p_0)$ and $f(\cdot,p)$ over $p \in {\cal P}$. }
\vspace{0.05in}

%For any distribution function with density $g_p$, one can easily choose $\zeta_{1n}$ and $\zeta_{2n}$, (or, $\alpha_{1n}$ and $\alpha_{2n}$) such that $(C1)~(i)$ holds. Moreover,
If $g_p$ is bounded away from zero and infinity, it is easy to see that $(C3)$ holds for any choice
of $\alpha_{1n}$ and $\alpha_{2n}$. In fact, $(C3)$ holds if $g_p$ remains bounded away from zero on any bounded interval inside its support. %Therefore, the conditions in $(C3)$ usually hold in practice.
Also note that the Sheather-Jones bandwidth (see Sheather and Jones, 1991) that we use for kernel density estimation is of the order $O_P(n^{-1/5})$. %This addresses condition $(C2)$.

%Since ${\bf b}=\muvec$ and ${\bf A}{\bf A}^{'}=\sigmat^{-1}$,Note that
The quantities ${\hat {\bf b}}$ and ${\hat {\bf A}}$ can be obtained directly from ${\hat \muvec}$ and ${\hat \sigmat}$, the estimates of ${\muvec}$ and ${\sigmat}$. Affine equivariant estimates of ${\muvec}$ and ${\sigmat}$ (e.g., usual moment based estimates or minimum covariance determinant (MCD) estimates) can be used for this purpose. Assuming the existence of second order moments, the sample mean converges to $\muvec$ and the sample dispersion matrix converges to constant multiple of $\sigmat$ at $\sqrt{n}$ rate as stated in $(C1)$.
% This constant depends on the underlying distribution (it is unity for some distributions).
%These convergences are at $\sqrt{n}$ rate as stated in $(C1)$.
Under appropriate regularity conditions, we have $\sqrt{n}$ convergence for MCD estimates as well (see Cator and Lopuha\"{a}, 2012).

\subsection{Estimation of ${\bf b}$ and ${\bf A}$}

We can continue to use ${\hat \muvec}$ (the moment based estimate, or the MCD estimate) as an estimate of ${\bf b}$. However, to construct ${\hat {\bf A}}$, one has to compute the square root of ${\hat \sigmat}$. Note that the symmetric square root of ${\hat \sigmat}$ is not affine eqivariant. We construct an affine equivariant square root of ${\hat \sigmat}$ using the transformation re-transformation technique (see, e.g., Chakraborty and Chaudhuri, 1996).
Consider a subset $\alpha=\{i_1,i_2,\ldots,i_{d+1}\}$ of the set $\{1,2,\ldots,n\}$. Use it to construct a $d \times d$ matrix ${\mathbb X}(\alpha)=[(\xvec_{i_1}-\xvec_{i_{d+1}})~ (\xvec_{i_2}-\xvec_{i_{d+1}})~\ldots~(\xvec_{i_d}-\xvec_{i_{d+1}})]$, and compute ${\mathbb Z}(\alpha)={\mathbb X}{(\alpha)}^{T}{\hat \sigmat}^{-1}~{\mathbb X}{(\alpha)}$. Now, find an $\alpha_0$ that makes ${\mathbb Z}(\alpha)$ close to a constant multiple of ${\bf I}_d$ (the $d \times d$ identity matrix).
For practical purposes, one may choose an $\alpha_0$ that maximizes the ratio of the determinant of ${\mathbb Z}(\alpha)$ to the trace of ${\mathbb Z}(\alpha)$.
%Under appropriate regularity conditions, Chakraborty (2001) proved that ${\mathbb X}(\alpha_0){\mathbb X}(\alpha_0)^{T}$ converges in probability to a scalar multiple of $\sigmat$, whenever ${\hat \sigmat}$ is consistent (up to a scalar multiple).
%\footnote{\color{red} Is this statement relevant here? \color{blue} Not sure. I couldn't find $\sqrt{n}$ consistency of ${\hat A}$, but we need to justify (C1).}
However, finding the actual maximizer $\alpha_0$ is computationally difficult.
%hard problem, especially when the sample size and the dimension are large. To reduce this computing cost, for our numerical work
So, following Chakraborty and Chaudhuri (1996), we choose an $\alpha$ which makes the ratio is close to $1$ ($\geq 0.99$).

\begin{figure}[h]
\vspace{-0.2in}
\begin{center}
\begin{tabular}{c c c}
% \hspace{-0.2in}
\vspace{-.1 in}
\includegraphics[width=5.50cm]{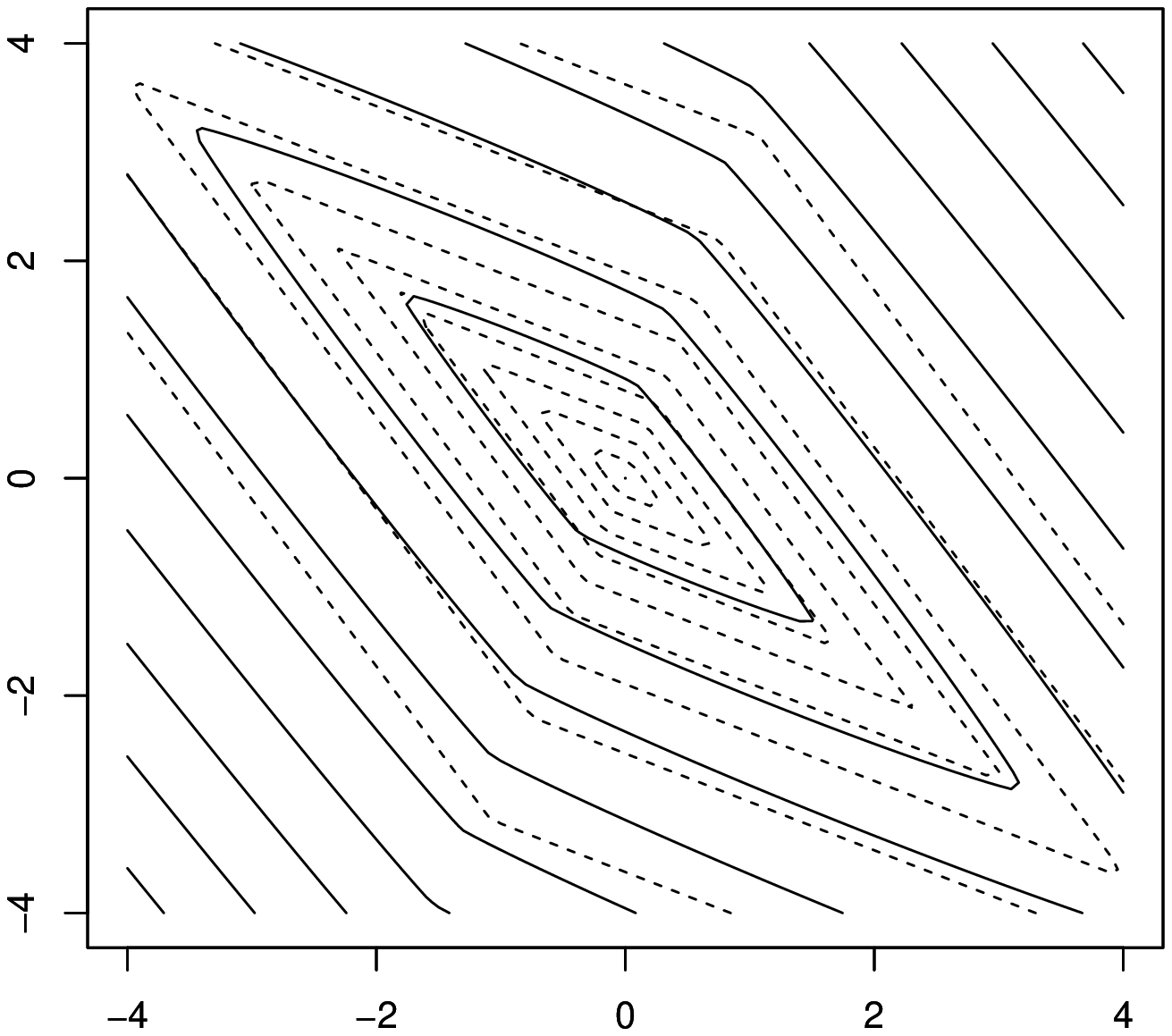}&
\includegraphics[width=5.50cm]{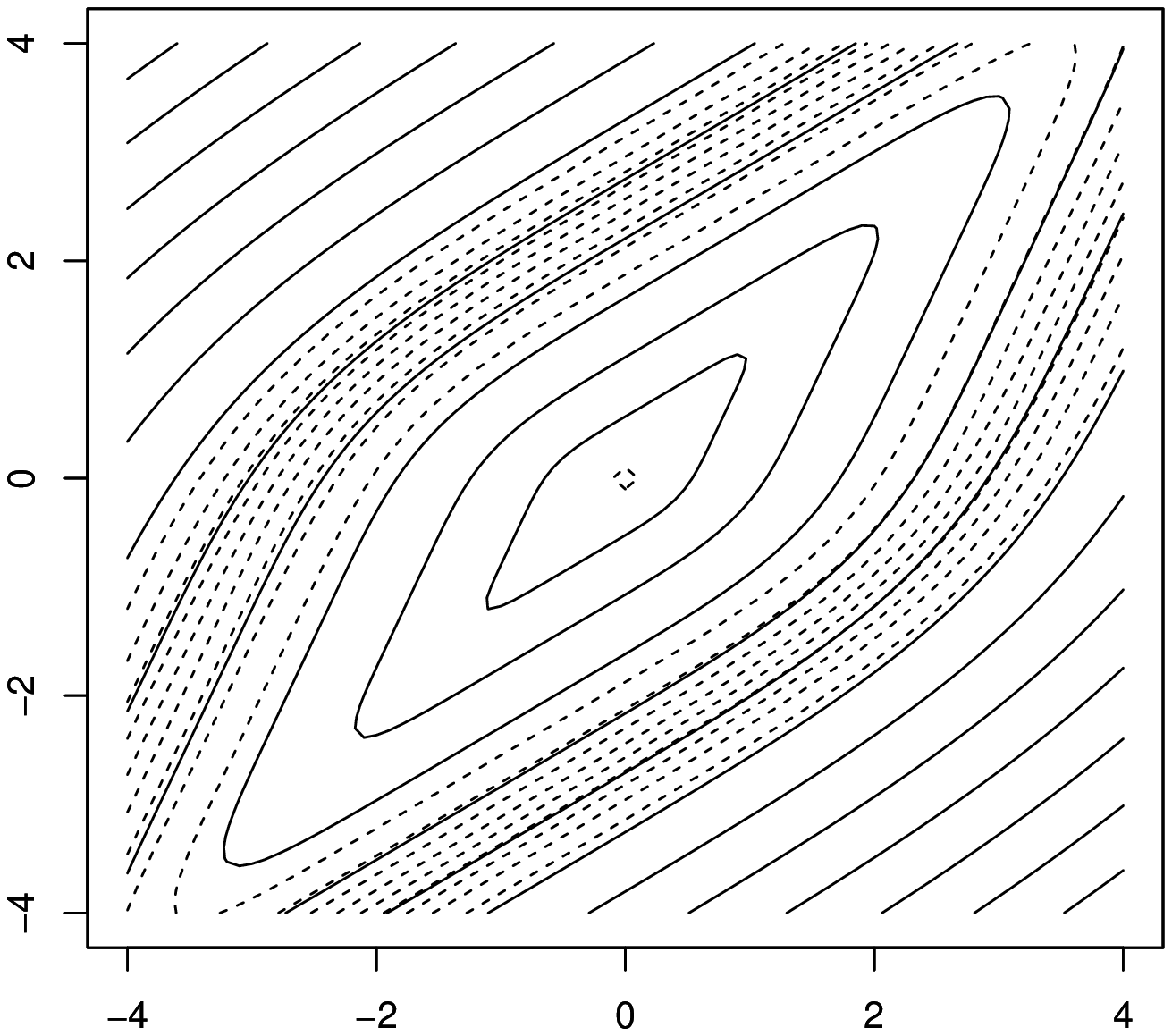}\\
\end{tabular}
\vspace{-0.1in}
\caption{Density contours ({\tt dotted curves}) and estimated L$_p$D contours ({\tt bold curves}).}
\end{center}
\vspace{-0.2in}
\end{figure}

The matrix ${\mathbb X}(\alpha_0)$ can now be considered as a square root of ${\hat \sigmat}$ (upto a scalar multiple). The use of ${\hat {\bf A}}={\mathbb X}^{-1}(\alpha_0)$ makes $\hat \delta_{p}(\xvec)$ affine invariant.
To study the empirical performance of this estimation method, we generated 400 observations from the density $f_1(x_1,x_2) \propto \exp\{-(|x_1|+{|x_2|}/{0.3})\}$ (which is $l_1$-symmetric), and then rotated them by an angle of $3\pi/4$. The left panel in Figure 2 shows that the estimated L$_p$D contours (solid curves) closely matches the corresponding density contours (dotted curves). We carried out a second experiment with observations from the density $f_2(x_1,x_2) \propto \exp\{-(|x_1|^5+{|x_2|^5}/{0.3})\}$ (which is $l_5$-symmetric), which were further rotated by an angle of $\pi/4$. In this example also, the estimation method worked quite well (see the right panel in Figure 2).
% observed the same when we

\section{Classification with L$_p$ Depth}

In this section, we first study the maximum L$_p$D classifier. This classifier works well when the competing classes have the same prior, and they differ only in their locations. However, if the classes have unequal priors and/or they differ only in their scatters and shapes, the maximum depth classifier may fail to have satisfactory performance. To cope with such cases, we develop a modified classifier in Section 3.2.

\subsection{Maximum L$_p$ depth classifier}

The maximum L$_p$D classifier classifies an observation to the class with respect to which it has the maximum depth. To construct the empirical version of L$_p$D, we estimate $p$ and $\delta_p(\xvec)$ from the data (as discussed in Section 2). Suppose there are $J$ competing classes with densities $f_1, \ldots, f_J$, and for the $j$-th class, these estimates are denoted by $\hat p_j$ and ${\hat \delta}_{\hat p_j,j}(\xvec)$, respectively, for $1 \leq j \leq J$. Then, the maximum L$_p$D classifier classifies an observation $\xvec$ to the $j$-th class if ${\hat \delta}_{\hat p_j,j}(\xvec) > {\hat \delta}_{\hat p_i,i}(\xvec)$ for all $i \neq j$.
However, this classifier is mainly used when the competing classes differ only in their locations, and in such cases it is more appealing to use a common value of $p$ for all classes. If $\xvec_{j1}, \xvec_{j2}, \ldots, \xvec_{jn_j}$ are observations from the $j$-th ($1 \leq j \leq J$) class, this common value of $p$ can be estimated by maximizing the joint log-likelihood function $\sum_{j=1}^{J} \varphi_{p} (\xvec_{j1}, \xvec_{j2}, \ldots, \xvec_{jn_j})$ over ${\cal P}$. The resulting maximum L$_p$D classifier is given by ${\sf d}_{1}(\xvec) = \arg \min_{1 \leq j \leq J} {\hat \delta}_{\hat p, j}(\xvec),$ where $\hat p$ denotes the common estimated value of $p$.

\vspace{0.05in}
{\bf Theorem 2:}
{\it Assume the density functions $f_1,\ldots,f_J$ to be unimodal, and $f_j$ is of the form $f_{j}(\xvec,p_0) = \psi(\|{\bf A}(\xvec-{\bf b}_j)\|_{p_{0}})$ for some $p_0 \in {\cal P}$ ($1 \leq j \leq J$). If the prior probabilities of the competing classes are equal, under conditions $(C1)$ - $(C3)$ (stated in Theorem 1), the misclassification rate of ${\sf d}_{1}(\xvec)$ converges to the Bayes risk as $min\{n_1,\ldots,n_J\} \rightarrow \infty$.}
\vspace{0.05in}

\subsection{Generalized L$_p$ depth classifier}

In practice, the prior probabilities of different classes may not be equal, and the class distributions can also differ in their scatters and shapes. In such cases, maximum depth classifiers may have higher misclassification probabilities (see, e.g., Dutta and Ghosh, 2012). We now construct the generalized L$_p$D classifier to cope with such situations. Suppose that there are $J$ competing classes, and  $f_{h_j}(\xvec, \hat p_j)$ is the density estimate for the $j$-th class ($1 \leq j \leq J$). In a two-class problem, this generalized L$_p$D classifier is given by
\begin{equation*}
{\sf d}_2(\xvec)= \left\{ \begin{array}{ll} 1 & ~\mbox{if}~~ \hat f_{h_1}(\xvec, \hat p_1)/\hat f_{h_2}(\xvec, \hat p_2) > k,\\
2 &~\mbox{otherwise}. \end{array} \right. \nonumber
\end{equation*}
%Under appropriate regularity conditions, $\hat f_{h_j}(\xvec, \hat p_j)$ converges to a scalar multiple of $f_{j}(\xvec, p_j)$ (see Lemma 3 in the Appendix). So, instead of choosing $k=\pi_2/\pi_1$ (where $\pi_1$ and $\pi_2$ are the prior probabilities of the first class and the second class,
%respectively), it is better to
where $k$ is chosen by minimizing the leave-one-out cross-validation estimate of the misclassification probability. %(see also Ghosh and Chaudhuri, 2005).
For more than two classes, we use the pairwise classification approach followed by the method of majority voting.
%For computation of $\hat f_{h_j}(\xvec, \hat p_j)$, we have used the univariate Sheather-Jones (1991) bandwidth as before.

\vspace{0.05in}
{\bf Theorem 3:} {\it Suppose that for all $1 \leq j \leq J$, the density $f_j$ is continuous, it has support over entire ${\mathbb R}^d$ and is of the form $f(\xvec, p_j^\circ) =\psi_j( \|{\bf A}_j(\xvec -{\bf b}_j)\|_{p_j^\circ})$ for some $p_j^\circ \in {\cal P}$. Also, assume that the optimal Bayes classifier has non-empty favorable regions for each of the $J$ classes. Then, under the conditions $(C1)$ - $(C3)$ (stated in Theorem 1), the misclassification rate of the generalized L$_p$D classifier ${\sf d}_2(\xvec)$ converges to the Bayes risk as $min \{n_1,\ldots,n_J\} \rightarrow \infty$.}
%\vspace{0.05in}

\subsection{Comparison with other depth based classifiers}
%Results from the analysis of simulated data sets}

To compare the performance of our L$_p$D classifiers with existing classifiers based on MD, HD and PD, we carried out some simulations. To keep our examples simple, here we used two-class problems in two dimensions.
%For $j=1,2$, we considered $f_j(\xvec, p_j)=c_{p_j}|{\bf A}_j| \exp\{-\|{\bf A}_j(\xvec-{\bf b}_j)\|_{p_j}^{p_j}\}$, where $c_{p_j}$ was the normalizing constant.
For varying choices of ($p_1,{\bf b}_1,{\bf A}_1$) and ($p_2,{\bf b}_2,{\bf A}_2$), we had different types of classification problems (see Table 1). In each case, taking equal number of observations from two competing classes, we generated training and test sets of sizes 400 and 1000, respectively. Each experiment was repeated 200 times, and the average test set misclassification rates of different classifiers were computed over these 200 trials. Note that our proposed method needs ${\cal P}$ to be specified. We observed that for higher values of $p$, $l_p$ contours do not change much (see, e.g., Figure 1, where $l_5$ contours almost look like $l_{\infty}$ contours). So, throughout this article, we used  ${\cal P}=\{2^{(i-1)/2}; ~ i=1,2,\ldots,10\}$ for our numerical work.

Following Ghosh and Hall (2008), we computed regret functions (i.e., difference between the misclassification rate of a classifier and the Bayes risk) for different classifiers. Table 1 shows the corresponding regret ratios for classifiers based on HD, PD and MD. Regret ratio ($\eta_t$) of the classifier $t$ is given by ratio of the regret of that classifier and that of the L$_p$D classifier. Clearly, $\eta_t <1(>1)$ implies that the classifier is better (worse) than the L$_p$D classifier, and the deviation from $1$ gives an idea of how better (worse) it is.  Since HD and PD classifiers are robust, here we used robust version of MD based on MCD estimates. Following Hubert and van Driessen (2004), 75\% observations were used to compute these estimates. These MCD estimates were also used as ${\hat \muvec}$ and ${\hat \sigmat}$ to compute ${\hat {\bf A}}$, ${\hat \bvec}$ and hence the empirical versions of L$_p$ depths.  For the location problems, we used maximum depth classifiers based on different notions on depth. In other examples, generalized depth based classifiers were used. In this section, we used equal prior probabilities for two competing classes.

In all these examples, L$_p$D classifiers outperformed the classifiers based on HD and PD. The classifiers based on HD had the worst performance in all cases, especially in location problems. The discrete nature of the empirical version of HD affected its performance, and it often had issues with ties and zero depth (also see Ghosh and Chaudhuri (2005)). As expected, the classifiers based on MD had a slight edge over the L$_p$D classifiers when both of the competing classes were $l_2$-symmetric. But, in all other cases, they were outperformed by the L$_p$D classifiers.

\vspace{-0.05in}
\begin{table}[t]
\setlength{\tabcolsep}{.9mm}
\begin{center}
\caption{Regret ratios of other depth-based classifiers.}
{\footnotesize
\vspace{0.15in}
\begin{tabular}{|c|c|c|c|c|c|c|c|c|c|c|c|c|}\hline
Choice $\downarrow$ & \multicolumn{4}{c|}{ Difference in locations} &\multicolumn{4}{c|}{
Difference in scales} &\multicolumn{4}{c|}{Difference in shapes}\\ \hline
$(p_1,{\bf b}_1,{\bf A}_1)$& $(p_2,{\bf b}_2,{\bf A}_2)$ & HD & PD & MD & $(p_2,{\bf b}_2,{\bf A}_2)$ & HD & PD & MD & $(p_2,{\bf b}_2,{\bf A}_2)$ & HD & PD & MD\\ \hline
$(1, {\bf 0}, {\bf I}_2)$ &$(1, {\bf 1}, {\bf I}_2)$ &53.01 &1.54 &1.26 & $(1, {\bf 0}, \frac{1}{9}{\bf I}_2)$ &2.59 &1.13 &1.01 &$(2, {\bf 0}, {\bf I}_2)$ &2.52 &2.49 &1.75\\
$(2, {\bf 0}, {\bf I}_2)$ &$(2, {\bf 1}, {\bf I}_2)$ &30.19 &2.97 &0.96 & $(2, {\bf 0}, \frac{1}{9}{\bf I}_2)$ &5.88 &1.50 &0.89 &$(4, {\bf 0}, {\bf I}_2)$ &1.57 &1.40 &1.14\\
$(8, {\bf 0}, {\bf I}_2)$ &$(8, {\bf 1}, {\bf I}_2)$ &25.94 &6.31 &1.17 & $(8, {\bf 0}, \frac{1}{9}{\bf I}_2)$ &5.84 &3.93 &2.06 &$(1, {\bf 0}, {\bf I}_2)$ &14.62 &2.85 &1.67\\ \hline
\end{tabular}
}

\vspace{0.1in}
{\footnotesize ${\bf 0}=(0,0)^{T}$, $~{\bf 1}=(1,1)^{T}$, $~{\bf I}_2$ is the $2 \times 2$ identity matrix.}
\end{center}
\vspace{-0.25in}
\end{table}
%\vspace{-0.15in}

\section{Results from the Analysis of Benchmark Datasets}

We analyzed eight benchmark data sets for further assessment of the generalized L$_p$D classifier. The hemophilia data set was taken from Johnson and Wichern (1992). All other data sets were taken either from the UCI Machine Learning Repository ({http://archive.ics.uci.edu/ml/}) or from the CMU Datasets Archive ({http://lib.stat.cmu.edu/datasets/}). Descriptions of these data sets are available at these sources. In the case of  blood transfusion data set, following Li et al. (2012), we considered only one of the two linearly dependent variables for our analysis. For the synthetic data and the satellite image (satimage) data, the training and the test sets are well specified. In all other cases, we formed these sets by randomly partitioning the data in a way such that the proportion of different classes in training and test sets were as close as possible. In cases of hemophilia data and diabetes data, we used training samples of size 50 and 100, respectively. In all other cases, we divided the data set into two nearly equal halves to form the training and the test sets. In each case, this random partitioning was done 500 times. Average  misclassification rates of different classifiers were computed over these 500 test sets, and they are reported in Table 3 along with their corresponding standard errors. In cases of synthetic data and satimage data, when a classifier led to a test set misclassification rate $\Delta$, its standard error was computed as $\sqrt{\Delta(1-\Delta)/n_t}$, for $n_t$ being the size of the test set.

To facilitate comparison, along with the misclassification rates of different depth based classifiers, results are also reported for two parametric classifiers (linear discriminant analysis (LDA) and quadratic discriminant analysis (QDA)) and two nonparametric classifiers (kernel discriminant analysis (KDA) and nearest neighbor classifier ($k$-NN) with the smoothing parameter
chosen by the leave-one-out cross-validation method). In some of these data sets, the measurement variables were not of comparable units and scales. So, we used KDA and $k$-NN (see, e.g., Hastie, Tibshirani and Friedman, 2009; Duda, Hart and Stork, 2012) on the standardized data set, where the moment based estimate of the pooled dispersion matrix was used for standardization. Therefore, to keep our comparisons fair, instead of MCD estimates, here we used moment based estimates of location and scatter parameters for MD and L$_p$D classifiers as well. Since the classifier based on HD had the highest misclassification rates in almost all cases, we do not report them here. For all these benchmark data sets, sample proportions of different classes were used as their prior probabilities.

%\vspace{-0.05in}
\begin{table}[htp]
\setlength{\tabcolsep}{.6mm}
% \hspace{-1 in}
\begin{center}
\caption{Average misclassification rates (in \%) of different classifiers and their standard errors.}
\vspace{0.1in}
{\small
\vspace{0.025in}
% \begin{tabular}{|@{\hspace{.5mm}}c@{\hspace{.5mm}}|@{\hspace{.5mm}}c@{\hspace{
% .5mm}}|@{\hspace{.5mm}}c@{\hspace{.5mm}}|@{\hspace{.5mm}}c@{\hspace{.5mm}}|@{
% \hspace{.5mm}}c@{\hspace{.5mm}}|@{\hspace{.5mm}}c@{\hspace{.5mm}}|@{\hspace{.5mm
% }}c@{\hspace{.5mm}}|@{\hspace{.5mm}}c@{\hspace{.5mm}}|@{\hspace{.5mm}}c@{\hspace
% {.5mm}}|@{\hspace{.5mm}}c@{\hspace{.5mm}}|}
\begin{tabular}{|c|c|c|c|c|c|c|c|c|c|c|c|}
\hline
Data & $d$ & $J$ & Train & Test & {LDA} & QDA & $k$-NN & KDA & PD &
MD & L$_p$D\\ \hline
Synthetic$^{\dagger}$ & 2 & 2 &250 & 1000 & 10.80 (.98) &10.20 (.96) &11.70 (1.0) &11.00 (.99) &10.80 (.98) &10.30 (.96) &{09.40} (.92)\\
% &  &  & & & (0.98) & (0.96) & (1.02) & (0.99) & (0.98) & (0.96) &(0.92)\\
%Salmon & 2 & 2  &50 & 50 & 08.18(0.14)& { 07.66}(0.13)& 08.76(0.15) & 08.63(0.14)& 09.31(0.17)&{07.49}(0.13)& { 07.51}(0.13)\\
% &  &   &  & &(0.14) & {(0.13)} & (0.15) & (0.14) &  (0.17) &{(0.13)} & (0.13) \\
Hemophilia & 2 & 2 &50 & 25 & 15.22 (.27)& 15.47 (.26)& 15.79 (.30) &15.11 (.27) &17.36 (.30)& 14.33 (.29)& 13.99 (.29)\\
% &  &  & & &(0.27)& (0.26)& (0.30) & (0.27)  & (0.30)& (0.29)& (0.29) \\
Blood Tran. & 3 & 2 & 374& 374& 22.90 (.03)&22.48 (.05) &21.39 (.06)&22.66 (.05)&22.27 (.08) &22.58 (.06) &{22.41} (.07)\\
% & &   & & &(0.03)& (0.05) & (0.06) & (0.05)& (0.08)  & (0.06) & (0.07)\\
Diabetes & 5 & 3 &100 & 45 & 10.46 (.18)&09.39 (.18)&10.04 (.18)&11.16 (.19)&10.38 (.23)&09.23 (.17)& {09.52} (.18)\\
% &  &  & & & (0.18)& (0.18)& (0.18)& (0.19)& (0.23)& (0.17) &(0.18)\\
Pima & 8 & 2 &384 &384 & 23.37 (.07) & 26.02 (.08)& 25.73 (.08)& 26.57 (.07)& 29.96 (.09)&25.47 (.08)&25.34 (.08)\\
% &  & & &  & (0.07)& (0.08) &  (0.08) &  (0.07)& (0.09) & (0.08)&(0.08)\\
%Vowel $^{\dagger}$& 10 &11 &528 & 462 & 55.63(2.31)&52.81(2.32)& 46.75(2.32)& 46.75(2.32) &  56.49(2.31)& 51.52(2.33) & 52.16(2.32) \\
%&  &  & & &(2.31) & (2.32)& (2.32)& (2.32) &  (2.31)&  (2.33) & (2.32)\\
Vehicle & 18 & 4 &423 &423 &22.49 (.07)  & 16.38 (.07)  & 21.84 (.08)  & 21.45 (.07)  & 42.64 (.11) & 16.29 (.07)  & 16.16 (.07)\\
% & &  & & &(0.07) & (0.07) & (0.08) &  (0.07) & (0.11) & (0.07) & (0.07) \\
Wisconsin & 30 & 2 &284 &285 & 04.71 (.05)  & 04.74 (.05)  & 09.36 (.07)  & 10.15 (.07)  & 07.45 (.06)   & 04.92 (.05)  & 04.63 (.05)\\
% &  &  &  & &(0.05) &  (0.05) &  (0.07) &  (0.07) &  (0.06)  &  (0.05) &  (0.05) \\
Satimage$^{\dagger}$ & 36 & 6 &4435 & 2000& 16.02 (.82) & 14.11 (.78) & 16.89 (.84) & 19.71 (.89) &19.30 (.88)  &{15.22} (.80) & {15.13} (.80)\\ \hline
%& &  & & &(0.82)& (0.78) & (0.84)& (0.89)& (0.88) & (0.80)&  (0.80)\\ \hline
\end{tabular}\\
}
\vspace{.1in}
$^{\dagger}${\footnotesize Data sets with specific training and test sets.}
%$^{\ast}${\footnotesize Lowest misclassification rate.}
\end{center}
\vspace{-0.2in}
\end{table}

The overall performance of the generalized L$_p$D classifier was fairly satisfactory (see Table 2). Except for the blood transfusion data, in all other data sets, it had lower misclassification rates than the PD classifier, and in most of the cases, the difference between their misclassification rates was found to be statistically significant at 5\% level when the usual large sample test was used for testing the equality of proportions. In 7 out of 8 data sets, it had lower misclassification rates than the MD classifier as well. It performed better than LDA in all cases barring the Pima Indian data and better than QDA in 6 out of 8 data sets. In cases of haemophilia data and Pima Indian data, its performance was significantly better than QDA. In several data sets, the generalized L$_p$D and MD classifiers performed better than KDA and $k$-NN classifiers as well. In high-dimensional data sets, especially in vehicle data and Wisconsin breast cancer (diagnostic) data, while nonparmetric classifiers like KDA and $k$-NN had poor performance due to data sparsity, generalized MD and L$_p$D classifiers were not affected much by curse of dimensionality. In these data sets, these two depth based classifiers yielded substantially lower misclassification rates.

\begin{figure}[h]
\begin{center}
\includegraphics[height=5.750cm,width=12.0cm]{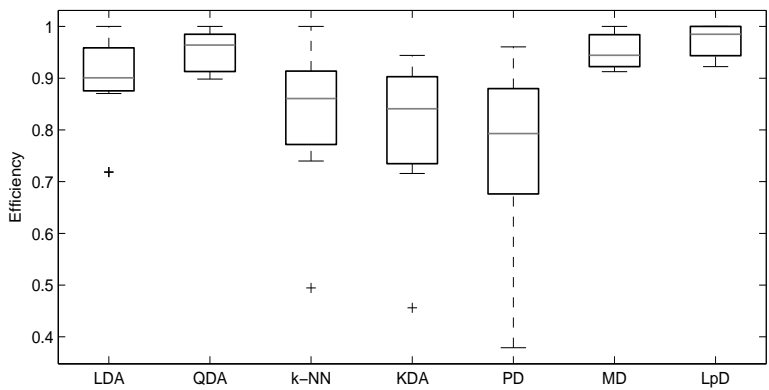}
\vspace{-0.1in}
\caption{Overall performance of different classifiers on benchmark data sets.}
\end{center}
\vspace{-0.15in}
\end{figure}

For visual comparison of the overall performance among different classifiers, we computed their efficiencies for different data sets, and they are presented using boxplots in Figure 3. In a particular data set, the efficiency of a classifier $t$ is given by $e_t =\Delta_0/\Delta_t$, where $\Delta_t$ is the misclassification rate of the classifier $t$ and $\Delta_0 =\min_t \Delta_t$. Clearly, $e_t=1$ for the best classifier, and $e_t<1$ for all other classifiers. Smaller values of $e_t$ indicates the lack of efficiency of the classifier $t$. Figure 3 clearly shows the superiority of the L$_p$D classifier over all other classifiers considered here.

\section*{Appendix: Proofs and Mathematical Details}

\noindent
{\bf Lemma 1}: If the density $f(\xvec,p)$ is of the form $\psi(\| {\bf A}(\xvec -{\bf b})\|_p)$ for some $p>0$ and a continuous function  $\psi: {\mathbb R}_+ \rightarrow {\mathbb R}_+$, then it can be expressed as
$$f(\xvec, p) = |{\bf A}|~ \frac{p^{d-1}\Gamma (d/p)}{2^d \{\Gamma (1/p)\}^d} \frac{g_p(\delta_p(\xvec))\delta_p(\xvec)^{d+1}}{[1-\delta_{p}(\xvec)]^{d-1}}~\mbox{for}~0<\delta_p(\xvec)<1,$$
%\vspace{-0.05in}
\noindent
where $\delta_p(\xvec)=[1+\|{\bf A}(\xvec-{\bf b})\|_p]^{-1}$, and $g_p$ is the density function of $\delta_p(\Xvec)$ when $\Xvec \sim f(\cdot,p)$.

\vspace{0.05in}
\noindent
{\bf Proof of Lemma 1}: Define $\Yvec = {\bf A}(\Xvec -{\bf b})$. If $f_0$ denotes the density of $\Yvec$, it is easy to see that $f_0(\yvec)=|{\bf A}|^{-1}\psi(\|\yvec\|_p)$. Now, following the proof of Lemma 1.4 in Fang, Kotz and Ng (1989), one can show that for any non-negative measurable function $\xi$, we have $$ \int_{\mathbb{R}^d} \xi(\| \yvec \|_p)~ d\yvec = \frac{2^d\{\Gamma
(1/p)\}^d}{p^d\Gamma (d/p)} \int_0^\infty \xi(u^{1/p}) u^{d/p-1} du.$$ So, for any non-negative measurable function $\phi$, defining $\xi=\phi.\psi$ we get
\vspace{-0.1in}
\begin{eqnarray}
\hspace{-0.2in}
E[\phi(\|\Yvec\|_p)] &=& |{\bf A}|^{-1} \int \phi(\|\yvec\|_p) \psi(\| \yvec \|_p)~ d\yvec
= |{\bf A}|^{-1}\frac{2^d \{\Gamma (1/p)\}^d}{p^{d}\Gamma (d/p)} \int_0^\infty \phi(u^{1/p})~\psi(u^{1/p})~u^{d/p-1}~du \nonumber \\
&=& |{\bf A}|^{-1} \frac{2^d \{\Gamma (1/p)\}^d}{p^{d-1}\Gamma (d/p)} \int_0^\infty \phi(r)~\psi(r)~r^{d-1}~dr ~(\mbox{letting} ~r = u^{1/p}). \nonumber
\end{eqnarray}

\vspace{-0.1in}
\noindent
Therefore, the density of $r_p(\Xvec)=\|{\bf A}(\Xvec-{\bf b})\|_p$ is the form
${\tau_p(r)=\disp |{\bf A}|^{-1}\frac{2^d \{\Gamma (1/p)\}^d}{p^{d-1}\Gamma (d/p)} ~\psi(r)~r^{d-1}}$, which implies %\linebreak
$f(\xvec,p)=\psi(r_p(\xvec))$ $= |{\bf A}|~\disp \frac{p^{d-1}\Gamma (d/p)}{2^d \{\Gamma (1/p)\}^d} ~\disp\frac{\tau_p(r_p(\xvec))}{r_p(\xvec)^{d-1}}.$
Recall that $\delta_p(\xvec)=[1+r_p(\xvec)]^{-1}$, and using the change of variables formula, we have the final expression for the density $f(\xvec,p)$ in terms of $\delta_p(\xvec)$ and its density $g_p(\cdot)$. \hfill $\Box$

\vspace{0.1in}
\noindent{\bf Corollary 1:} Define ${\tilde \delta}_p(\xvec)=[1+a_0r_p(\xvec)]^{-1}$, where $a_0$ is a positive constant, and let $g_p^{(a)}(\cdot)$ denote the density of ${\tilde \delta}_p(\Xvec)$. If the density $f(\xvec,p)$ is of the form $\psi(\| {\bf A}(\xvec -{\bf b})\|_p)$ for some $p>0$ and a continuous function $\psi: {\mathbb R}_+ \rightarrow {\mathbb R}_+$, it can be expressed as
$$f(\xvec, p) = a_0^d ~|{\bf A}|~ \frac{p^{d-1}\Gamma (d/p)}{2^d \{\Gamma (1/p)\}^d} \frac{g_p^{(a)}({\tilde \delta}_p(\xvec)){\tilde \delta}_p(\xvec)^{d+1}}{[1-{\tilde \delta}_{p}(\xvec)]^{d-1}}~\mbox{for}~0<{\tilde \delta}_p(\xvec)<1,$$

%\vspace{0.05in}
\noindent
{\bf Proof of Corollary 1}: If we define ${\tilde r}_p(\xvec) = a_0 r_p(\xvec)$ and if $\tau^{(a)}_p(\cdot)$ denotes the density of ${\tilde r}_p(\Xvec)$, it is easy to check that
$\disp\frac{\tau_p(r_p(\xvec))}{r_p(\xvec)^{d-1}}=a_0^d \disp\frac{\tau_p^{(a)}({\tilde r}_p(\xvec))}{{\tilde r}_p(\xvec)^{d-1}}$. Now from the proof of Lemma 1, it follows that
$f(\xvec,p)= a_0^d~ |{\bf A}|~\disp \frac{p^{d-1}\Gamma (d/p)}{2^d \{\Gamma (1/p)\}^d} ~\disp\frac{\tau_p^{(a)}({\tilde r}_p(\xvec))}{{\tilde r}_p(\xvec)^{d-1}}.$ Since ${\tilde \delta}_p(\xvec)=[1+{\tilde r}_p(\xvec)]^{-1}$, the result is obtained by using the change of variables formula. \hfill $\Box$

\vspace{0.1in}
\noindent
{\bf Lemma 2}: Let $f$ be $l_{p_0}$-symmetric for some $p_0\ge 1$. Define ${\tilde \delta}_p(\xvec)$ and $g_p^{(a)}$ as in Corollary 1.\\
(i) If $\hat{{\bf b}}$ and ${\hat {\bf A}}$ satisfy assumption $(C1)$, for any $p\ge 1$, $\sup_{\xvec \in \mathbb{R}^d} |{\hat \delta}_p(\xvec)-{\tilde \delta}_p(\xvec)|=O_P(n^{-1/2})$. \\
(ii) Further, if the density $g_p$ and the bandwidth $h$ associated with the kernel density estimator ${\hat g}_{p,h}(\cdot)$ satisfies assumption $(C2)$, then $\sup_{\xvec \in \mathbb{R}^d} |{\hat g}_{p,h}(\hat \delta_p(\xvec)) - g_p^{(a)}({\tilde\delta}_p(\xvec))| \stackrel{P}{\rightarrow} 0~\mbox{as}~n \rightarrow \infty$.
%of the order $O(n^{-1/5})$,
%\footnote{\color{red} define $\delta_p(\xvec)$ as $[1+r_p(\xvec)]^{-1}$?}
%\footnote{\color{red} results are in P or a.s.?}

\vspace{0.05in}
\noindent
{\bf Proof of Lemma 2}: (i) For any matrix $\Avec$, let $\|\Avec\|_F=[trace(\Avec^{T}\Avec)]^{1/2}$ denote its Frobenius norm and $\|\Avec\|_p=\sup_{\xvec} \{\|A\xvec\|_p/\|\xvec\|_p\}$ denote its $p$-th norm. For any $p \geq 1$, $\|\Avec\|_p$ is an induced norm that satisfies $\|\Avec \xvec\|_p \leq \|\Avec\|_p \|\xvec\|_p$ for all $\xvec \in {\mathbb R}^d$. %Note that $\|\Avec\|_2 \leq \|\Avec\|_F$, and using this we get $\|\Avec \xvec\|_p \leq \|\Avec\|_F \|\xvec\|_p$.
%$\|\xvec\|_p \leq \|\xvec\|_1$ for any $p \geq 1$.
%Combining these inequalities,
% and $|\hat{d}_p(\xvec) - {d}_p(\xvec)| \leq \disp \frac{|\hat{r}_p(\xvec) - {r}_p(\xvec)|}{[\hat{r}_p(\xvec)][{r}_p(\xvec)]}$.
In finite dimension, since all norms are equivalent, we have $\|\Avec \xvec\|_p \leq C_p\|\Avec\|_F \|\xvec\|_p$ for some positive constant $C_p$. This also implies that $\|\xvec\|_p \le C_p\|\Avec^{-1}\|_F \|\Avec\xvec\|_p$, or $\|\Avec\xvec\|_p \ge \|\xvec\|_p/C_p\|\Avec^{-1}\|_F$.

Let us divide ${\mathbb R}^{d}$ into two disjoint regions: ${\cal R}_1=\{\xvec : \|\xvec\| \le M\}$ and ${\cal R}_2=\{\xvec : \|\xvec\| > M\}$, where $M$ is a large positive constant. Note that $|\hat{\delta}_p(\xvec) - \tilde{\delta}_p(\xvec)| = \disp \frac{|\hat{r}_p(\xvec) - a_0{r}_p(\xvec)|}{[1+\hat{r}_p(\xvec)][1+a_0{r}_p(\xvec)]} \leq |\hat{r}_p(\xvec) - a_0{r}_p(\xvec)|$. Using the triangle inequality, we obtain
\vspace{-0.1in}
$$ \sqrt{n} \sup_{\xvec \in {\cal R}_1}|\hat{\delta}_p(\xvec) - \tilde{\delta}_p(\xvec)|
 \leq \sqrt{n} \sup_{\xvec \in {\cal R}_1} |\hat{r}_p(\xvec) - a_0{r}_p(\xvec)| \leq \|\hat{\Avec}\|_F \sqrt{n} \|\hat{\bvec}-\bvec\|_p +
 \sqrt{n} \|\hat{\Avec}-{a_0\Avec}\|_{F} \sup_{\xvec \in {\cal R}_1} \|\xvec-\bvec\|_p.
\vspace{-0.1in} $$
Since $\sup_{\xvec \in {\cal R}_1} \|\xvec-\bvec\|_p \le M+\|\bvec\|$,  under the condition ($C1$), $\sup_{\xvec \in {\cal R}_1} \sqrt{n} |\hat{\delta}_p(\xvec) - \tilde{\delta}_p(\xvec)| =O_p(1)$.
Also note that $|\hat{\delta}_p(\xvec) - \tilde{\delta}_p(\xvec)| \leq \disp \frac{|\hat{r}_p(\xvec) - a_0{r}_p(\xvec)|}{[\hat{r}_p(\xvec)][a_0{r}_p(\xvec)]}$
and this implies that
\vspace{-0.1in}
$$\sqrt{n} |\hat{\delta}_p(\xvec) - \tilde{\delta}_p(\xvec)|  \leq \disp \frac{ \|\hat{\Avec}\|_F \sqrt{n} \|\hat{\bvec}-\bvec\|_p}{(\|\hat{\Avec}\xvec\|_p - \|\hat{\Avec} \hat{\bvec}\|_p)(\|{a_0\Avec}\xvec\|_p - \|{a_0\Avec}{\bvec\|}_p)} + \frac{\sqrt{n} \|\hat{\Avec}-{(a_0\Avec)}\|_{F}}{(\|\hat{\Avec}\xvec\|_p - \|\hat{\Avec} \hat{\bvec}\|_p) a_0 \|{\Avec}\|_F}.$$
{Therefore}, $\sqrt{n} \sup_{\xvec \in {\cal R}_2}|\hat{\delta}_p(\xvec) - \tilde{\delta}_p(\xvec)|$
$$\leq \disp \frac{ \|\hat{\Avec}\|_F \sqrt{n} \|\hat{\bvec}-\bvec\|_p}{[(M/\|\hat{\Avec}^{-1}\|_F )- \|\hat{\Avec} \hat{\bvec}\|_p][(M/\|{C_pa_0\Avec}^{-1}\|_F) - \|{a_0\Avec}{\bvec\|}_p]} + \frac{\sqrt{n} \|\hat{\Avec}-{(a_0\Avec)}\|_{F}}{[M/\|\hat{\Avec}^{-1}\|_F) - \|\hat{\Avec} \hat{\bvec}\|_p] a_0 \|{\Avec}\|_F}.$$
%One can check that for
In each of the two terms on the right side, the numerator is $O_p(1)$, while the denominator coverges to a positive constant. So, we have
$\sqrt{n} \sup_{\xvec \in {\cal R}_2}|\hat{\delta}_p(\xvec) - \tilde{\delta}_p(\xvec)|=O_p(1)$. Now, combining the results on ${\cal R}_1$ and
${\cal R}_2$, the first part of the lemma is proved.

(ii) Note that
$|{\hat g}_{p,h}(\hat \delta_p(\xvec)) - g_p^{(a)}(\tilde{\delta}_p(\xvec))| \leq |\hat g_{p,h}(\hat \delta_p(\xvec)) - \hat g_{p,h}(\tilde{\delta}_p(\xvec))| + |\hat g_{p,h}(\tilde{\delta}_p(\xvec)) - g_p^{(a)}(\tilde{\delta}_p(\xvec))|.$
An application of the triangle inequality implies that
$$|\hat g_{p,h}(\hat \delta_p(\xvec)) - \hat g_{p,h}(\tilde{\delta}_p(\xvec))| \leq \frac{1}{nh} \sum_{i=1}^n \left | K \left [\frac{\hat \delta_p(\xvec)- \hat \delta_p(\xvec_i)}{h} \right ] \right. -
\left. K \left [\frac{\tilde{\delta}_p(\xvec)-\hat \delta_p(\xvec_i)}{h} \right ] \right |, $$
where $K$ is the kernel function associated with the density estimate. Using the mean value theorem and assuming $C_K = \sup_x |K^{\prime}(x)| < \infty$ (here we use the Gaussian kernel, which has bounded first derivative), we have
$\sup_{\xvec \in \mathbb{R}^d} |\hat g_{p,h}(\hat \delta_p(\xvec)) - \hat
g_{p,h}(\tilde{\delta}_p(\xvec))| \leq C_K {\sup_{\xvec \in \mathbb{R}^d}
|\hat \delta_p(\xvec) - \tilde{\delta}_p(\xvec)|}/{h^2}$. So, if $h$ satisfies the condition ($C2$),
%for any $\eta > \delta/2$ (note that $\eta=1/20$ if we use $h$ of the order $O(n^{-1/5})$),
$~\sup_{\xvec \in \mathbb{R}^d} |\hat g_{p,h}(\hat \delta_p(\xvec)) - \hat g_{p,h}(\tilde{\delta}_p(\xvec))| \stackrel{P}{\rightarrow} 0$ as $n \rightarrow \infty$.
%\noindent

To prove the convergence of  $\sup_{\xvec \in \mathbb{R}^d} |\hat g_{p,h}(\tilde{\delta}_p(\xvec)) -g_p^{(a)}(\tilde{\delta}_p(\xvec))|$,
first note that
\vspace{-0.1in}
$$\sup_{\xvec \in \mathbb{R}^d} |\hat g_{p,h}(\tilde{\delta}_p(\xvec)) -
g_p^{(a)}(\tilde{\delta}_p(\xvec))| \leq \sup_{\xvec \in \mathbb{R}^d} |\hat g_{p,h}(\tilde \delta_p(\xvec)) - g_{p,h}^{*}(\tilde \delta_p(\xvec))| + \sup_{\xvec \in \mathbb{R}^d} |g_{p,h}^{*}(\tilde \delta_p(\xvec)) - g_p^{(a)}(\tilde \delta_p(\xvec))|,
\vspace{-0.1in}$$
where $g_{p,h}^{*}(\tilde \delta_p(\xvec)) = {(nh)}^{-1} \sum_{i=1}^n K[\{\tilde \delta_p(\xvec)- \tilde \delta_p(\xvec_i)\}/{h}]$.
So, using triangle inequality, one gets
\vspace{-0.1in}
$$\sup_{\xvec \in \mathbb{R}^d} |\hat g_{p,h}(\tilde \delta_p(\xvec)) - g_p^{(a)}(\tilde \delta_p(\xvec)) \le \sup_{\xvec \in \mathbb{R}^d} |g_{p,h}^{*}(\tilde \delta_p(\xvec)) - g_p^{(a)}(\tilde \delta_p(\xvec))| + C_K {\sup_{\xvec \in \mathbb{R}^d} |\hat \delta_p(\xvec) - \tilde \delta_p(\xvec)|}/{h^2}.
\vspace{-0.1in}$$
We have already proved the convergence of the second term on the right side to $0$. Since $g_p$ is uniformly continuous and $nh/\log(n) \rightarrow \infty$ as $n \rightarrow \infty$ (see assumption (C2)), the convergence of the first term and hence the result now follow from the uniform convergence property of the kernel density estimate $g_{p,h}^*$ (see, e.g., Silverman, 1998).
%\footnote{\color{red} need uniform continuity of $g_p$ \color{blue} Yes, we need it and it has been mentioned in (C2).}
%$$~~~~~~~~~~~~~~~~~~~~~~~~~~~~~~~~~~~~~~~~~~~~~~~~~~~\sup_{\xvec \in \mathbb{R}^d} |\hat g_{p,h}(\tilde \delta_p(\xvec)) - g_p^{(a)}(\tilde \delta_p(\xvec))|
%\stackrel{P}{\rightarrow} 0 ~\mbox{as}~ n \rightarrow \infty. ~~~~~~~~~~~~~~~~~~~~~~~~~~~~~~~~~~~~~~~~~~~~\hfill \Box$$

%\newpage
\vspace{0.1in}
\noindent
{\bf Proof of Theorem 1}:  Define $A_n=\{\xvec : \tilde \delta_{p}(\xvec) \in [\zeta_{1n}, \zeta_{2n}]\} \subset \mathbb{R}^d$. Using the mean value theorem on the logarithmic function, one gets
$$\sup_{\xvec \in A_n} |\ln \hat g_{p,h}(\hat \delta_p(\xvec)) - \ln g_p^{(a)}(\tilde \delta_p(\xvec))| \leq  \sup_{\xvec \in \mathbb{R}^d} |\hat g_{p,h}(\hat \delta_p(\xvec)) - g_p^{(a)}(\tilde \delta_p(\xvec))| \sup_{\xvec \in A_n} \left \{ \frac{1}{\xi_n(\xvec)} \right \}, $$
where $\xi_n(\xvec) = \lambda_n \hat g_{p,h}(\hat \delta_p(\xvec)) + (1 - \lambda_n) g_p^{(a)}(\tilde \delta_p(\xvec))$ for some $\lambda_n \in (0,1)$. Hence, $\xi_n(\xvec)$ converges to $g_p^{(a)}(\tilde \delta_p(\xvec))$ as $n \rightarrow \infty$ (follows from Lemma 2). %\footnote{\color{red} confusion with definition of $a_p$! \color{blue} Removed}
%Since the Sheather-Jones bandwidth is of the order $O_P(n^{-1/5})$,
From the proof of Lemma 2 and using a result on the rate of uniform consistency of the kernel density estimate (see Theorem B in Silverman, 1978, p. 181). Under (C2), we have $\sup_{\xvec \in \mathbb{R}^d} |\hat g_{p,h}(\hat \delta_p(\xvec)) -g_p^{(a)}(\tilde \delta_p(\xvec))| = O_P(n^{-1/2}h^{-2})$. Therefore, using $(C3)$, we get
$\sup_{\xvec \in A_n} |\ln \hat g_{p,h}(\hat \delta_p(\xvec)) - \ln g_p^{(a)}(\tilde \delta_p(\xvec))| \stackrel{P}{\rightarrow} 0~\mbox{as}~n \rightarrow \infty$.

Following similar arguments as above, one can show that
\vspace{-0.1in}$$ \sup_{\xvec \in A_n} |\ln \hat \delta_p(\xvec) - \ln (\tilde \delta_p(\xvec))| \leq \sup_{\xvec \in \mathbb{R}^d} | \hat \delta_p(\xvec) - (\tilde \delta_p(\xvec))| \sup_{\xvec \in A_n} \left \{\frac{1}{\gamma_n(\xvec)} \right \},$$
where $\gamma_n(\xvec)$ = $\lambda_n \hat \delta_p(\xvec) + (1 - \lambda_n) \tilde \delta_p(\xvec)$ for some $\lambda_n \in (0,1)$, and it converges to $\tilde \delta_p(\xvec)$. Since $\sup_{\xvec \in \mathbb{R}^d} | \hat \delta_p(\xvec) - (\tilde \delta_p(\xvec))|=O_P(n^{-1/2})$ and $n^{1/2}\zeta_{1n} \rightarrow \infty$ as $n\rightarrow \infty$,
%\footnote{\color{red} how? \color{blue} After the changes in assumptions, it is clear now}
we have
$\sup_{\xvec \in A_n} |\log \hat \delta_p(\xvec) - \log (\tilde{\delta}_p(\xvec))|$ $\stackrel{P}{\rightarrow} 0~\mbox{as}~n \rightarrow \infty$. Similarly, using the fact that $n^{1/2}(1-\zeta_{2n}) \rightarrow \infty$ as $n\rightarrow \infty$, under the given conditions, one can show that $\sup_{\xvec \in A_n} |\log (1-\hat \delta_p(\xvec)) - \log (1-{\tilde{\delta}}_p(\xvec)))|\stackrel{P}{\rightarrow} 0~\mbox{as}~n \rightarrow \infty$. Combining these results and using corollary 1, we get $\sup_{\xvec \in A_n}|\ln \hat f_n(\xvec, p) - \ln (a_0^{-d}f(\xvec, p))| \stackrel{P}{\rightarrow} 0$ as $n \rightarrow \infty.$
%\footnote{\color{red} how do we get $a_0^{-d}f(\xvec, p)$? \color{blue}Corollary 1}

Define the set $I_n=\{i: \Xvec_i \in A_n\}$ and note that
$\Bigl| \frac{1}{n} \sum_{i~\in I_n} \ln \hat f_n(\xvec_i, p) - E_{p_0} \{ \ln (a_0^{-d}f(\Xvec, p)) \} \Bigr|
\leq \sup_{\xvec \in A_n}|\ln \hat f_n(\xvec, p)$ - $\ln
(a_0^{-d}f(\xvec, p))|$ $+ \left | \frac{1}{n} \sum_{i~\in I_n} \ln
(a_0^{-d}f(\xvec_i, p)) - E_{p_0} \{ \ln (a_0^{-d}f(\Xvec, p)) \}
\right |,$
where $E_{p_0}$ denotes the expectation with respect to $f(\xvec, p_0)$. We have already proved the probability convergence of the first part on the right side to $0$. From the strong law of large numbers, we have
$\left | \frac{1}{n} \sum_{i~\in I_n} \ln
(a_0^{-d}f(\xvec_i, p))  - E_{p_0}\{ \ln (a_0^{-d}f(\Xvec, p)) I(\Xvec \in A_n) \} \right | \stackrel{a.s.}{\rightarrow} 0$ as $n \rightarrow \infty$.

Using monotone convergence theorem on the positive and negative parts of the integrand separately, we have $E_{p_0}\{ \ln (a_0^{-d}f(\Xvec, p)) I(\Xvec \in A_n) \} \rightarrow E_{p_0}\{ \ln (a_0^{-d}f(\Xvec, p)) \}~\mbox{as}~n \rightarrow \infty$.
%\footnote{\color{red} DCT needs $f$ to be bounded! \color{blue} MCT}
So, for all $p \geq 1$, we get
$\frac{1}{n} \sum_{i~\in~ I_n} \ln \hat f_n(\xvec_i, p) \stackrel{P}{\rightarrow} E_{p_0} \{ \ln (a_0^{-d}f(\Xvec, p)) \}=\ln(a_0^{-d}) + E_{p_0}\{ \ln f(\Xvec, p) \}$. One can notice that $E_{p_0}\{ \ln f(\Xvec, p) \}= E_{p_0}\{ \ln f(\Xvec, p_0) \} - KL(p_0,p)$, where $KL(p_0,p)$ denotes the Kullback-Leibler divergence between $f(\Xvec, p_0)$ and $f(\Xvec, p)$. So, maximization of $E_{p_0}\{ \ln f(\Xvec, p) \}$ is equivalent to minimization of $KL(p_0,p)$ over ${\cal P}$. Using Jensen's inequality on the logarithmic function, we also have $E_{p_0}\{ \ln f(\Xvec, p) \} < E _{p_0}\{ \ln f(\Xvec, p_0) \}$, or $KL(p_0,p)>KL(p_0,p_0)$ for all $p \neq p_0$. Since ${\cal P}$ is a finite set, the proof now follows from the uniqueness of the minimizer of $KL(p_0,p)$.  \hfill $\Box$
%maximizer $E_{p_0}\{ \ln f(\Xvec, p) \}$ or the
%Note that since $\hat f_n(\xvec, p)$ and
%$E_{p_0}\{ \ln f(\Xvec, p) \}$ both are continuous functions of $p$, this result holds even when ${\cal P}$ is not finite (see Lemma 5.10 in van der Vaart,~2000).

%\newpage
\vspace{0.1in}
\noindent
{\bf Proof of Theorem 2}: The misclassification rate of the classifier ${\sf d}_{1}$ is given by

\vspace{-0.2in}
$${ \Delta({\sf d}_{1}) = \frac{1}{J}\sum_{j=1}^{J} \int \theta_j(\xvec,{\hat
p_n}) f_j(\xvec,p_0) ~d\xvec ,}$$
\vspace{-0.2in}

\vspace{-0.1in}
\noindent
where $\theta_j(\xvec,{\hat p_n})=P\{ \hat{\delta}_{{\hat p_n},j}(\xvec) < \hat{\delta}_{{\hat p_n},i}(\xvec)
~\mbox{for some}~~ i\neq j\}$. From the arguments given in the proof of Theorem 1, it follows that $\hat p_n \stackrel{P}{\rightarrow} p_0$. Using the continuous mapping theorem, one gets $\hat{\delta}_{{\hat p_n},j}(\xvec) \stackrel{P}{\rightarrow} \tilde {\delta}_{p_0,j}(\xvec)$ for all $1 \leq j \leq J$.
%, where $a_0$ is a constant that depends on $\psi$, $\sigmat$ and $p_0$.
So, if $\tilde {\delta}_{p_0,j}(\xvec) < \tilde {\delta}_{p_0,i}(\xvec)$ for all $i \neq j$ (or equivalently, $f_j(\xvec,p_0) > f_i(\xvec,p_0)$ for all $i \neq j$), $\theta_j(\xvec,{\hat p}_n)$ converges to $0$, otherwise it converges to $1$. Therefore, using
the Dominated Convergence Theorem, we have the convergence of $\Delta({\sf d}_{1})$ to the Bayes risk. \hfill $\Box$

\vspace{0.1in}
\noindent
{\bf Lemma 3}: If $p_0 \in {\cal P}$, under $(C1)$ - $(C3)$, for any fixed $\xvec \in \mathbb{R}^d$,  $\hat f_h(\xvec, p_0) \stackrel{P}{\rightarrow} a_0^{-d}f(\xvec, p_0)$ as $n \rightarrow \infty$.

\vspace{0.1in}
\noindent
{\bf Proof of Lemma 3}: Recall that $\displaystyle \hat f_h(\xvec, p_{0}) = C_{\hat p_0, d} \frac{\hat g_{p_0,h}(\hat {\delta}_{p_0}(\xvec))(\hat {\delta}_{p_0}(\xvec))^{d+1}}{(1-\hat {\delta}_{p_0}(\xvec))^{d-1}},$ where $C_{p,d} = \disp \frac{p^{d-1}\Gamma (d/p)}{2^d \{\Gamma (1/p)\}^d}$. From the continuous mapping theorem and Theorem 1, we get
$(\hat {\delta}_{\hat p_n}(\xvec))^{d+1} \stackrel{P}{\rightarrow} (\tilde {\delta}_{p_0}(\xvec))^{d+1}$, $(1-\hat {\delta}_{\hat p_n}(\xvec))^{d-1} \stackrel{P}{\rightarrow} (1-\tilde {\delta}_{p_0} (\xvec))^{d-1} ~\mbox{for}~d > 1$ and $\hat g_{\hat p_0,h}( {\delta}_{p_0}(\xvec)) \stackrel{p}{\rightarrow} g_{p_0}^{(a)}({\delta}_{p_0}(\xvec))$. The result now follows from Slutsky's lemma and Corollary 1. \hfill $\Box$

\vspace{0.1in}
\noindent
{\bf Proof of Theorem 3 }: For simplicity, we consider the case when $J=2$. For $J>2$, we use pairwise classification, and hence the result can be obtained by repeating the same argument for each of the $J(J-1)/2$ pairs of classes (see Tewari and Bartlett, 2007). %In a two class problem, 
Using Lemma 3, we get
$(\hat f_{h_1}(\xvec, p_{1}^{\circ}), \hat f_{h_2}(\xvec, p_{2}^{\circ})) \stackrel{P}{\rightarrow} (a_{1}^{-d}f(\xvec, p_1^{\circ}), a_{2}^{-d}f(\xvec, p_2^{\circ})),~\mbox{where}~a_{1}, a_{2}> 0$. Consider the classifier ${\sf d}_2^{p_1^{\circ},p_2^{\circ}}$ which is of the form:

\vspace{-0.225in}
\begin{equation*}
{\sf d}_2^{p_1^{\circ},p_2^{\circ}}(\xvec)= \left\{ \begin{array}{ll} 1 & ~\mbox{if}~~ \hat f_{h_1}(\xvec, p_1^{\circ})/\hat f_{h_2}(\xvec, p_2^{\circ}) > k,\\
2 &~\mbox{otherwise}. \end{array} \right. \nonumber
\end{equation*}
Here, $k$ is chosen by minimizing the leave-one-out cross-validation estimate of the misclassification probability. Note that for any fixed $k$, this cross-validation estimate is given by

\vspace{-0.175in}
$$\Delta^{CV}_{n,p_1^{\circ},p_2^{\circ}}(k) =  \sum_{i=1, j \neq i}^{2} \frac{\pi_i}{n_i}\sum_{l=1}^{n_{i}}  I \left[ \Bigl\{{\hat f_{h_i}^{-il}(\xvec_{il}, p_{i}^{\circ})} / {\hat f_{h_j}^{-il}(\xvec_{jl},p_{j}^{\circ})}\Bigr\} \leq k_i \right],$$
\vspace{-0.175in}

\vspace{-0.075in}
\noindent
where $n=(n_1, n_2)$, $k_1 = k$, $k_2 = 1/k$, and ${\hat f}_{h_j}^{-il}$ denotes the leave-one-out density estimate with $\xvec_{il}$ being left out as a training data point. Let $k_{n}$ denote the minimizer of $\Delta_{n,p_1^{\circ},p_2^{\circ}}^{CV}(k)$.

Define
\vspace{-0.275in}
$$\Delta_{p_1^{\circ},p_2^{\circ}}(k) = \sum_{i=1, j \neq i}^{2} \pi_i P \left[ \Bigl\{{a_{i}^{-d}f(\Xvec, p_{i}^{\circ})}/ {a_{j}^{-d}f(\Xvec, p_{j}^{\circ})} \Bigr\}\leq k_i ~\biggr|~ \Xvec \in i \mbox{-th class} \right], $$

\vspace{-0.075in}
\noindent
and $k_0 = \arg \min_k \Delta_{p_1^{\circ},p_2^{\circ}}(k)$. Under the stated assumptions, $k_0=a_{1}^{-d}\pi_2/a_{2}^{-d}\pi_1$ is a unique solution, and $0<k_0<\infty$. Note that $\Delta_{p_1^{\circ},p_2^{\circ}}(k_0)$ denotes the misclassification probability of the Bayes classifier ${\sf d}_B$. Consistency of the density estimates and using the same arguments as in Lemma 3 of Dutta and Ghosh (2012), one can show that $\sup_k |\Delta_{n,p_1^{\circ},p_2^{\circ}}^{CV}(k)- \Delta_{p_1^{\circ},p_2^{\circ}}(k)| \stackrel{P}{\rightarrow} 0$. This now implies that $k_{n} \stackrel{P}{\rightarrow} k_0$ as $\min\{n_1,n_2\} \rightarrow \infty$.
So, using the continuous mapping theorem and continuity of the underlying densities, we have ${\sf d}_2^{p_1^{\circ},p_2^{\circ}}(\xvec) \stackrel{P}{\rightarrow} {\sf d}_B(\xvec)$ for any fixed $\xvec \in \mathbb{R}^d$ not lying on the Bayes class boundary. %Therefore, $ I({\sf d}_{p_,p_2,{\nvec}}(\xvec) \neq i) \stackrel{P}{\rightarrow} I({\sf d}_B(\xvec) \neq i) $ for $i = 1, 2$.
Using the Dominated Convergence Theorem, we can show that the misclassification probability of the classifier ${\sf d}_2^{p_1^{\circ},p_2^{\circ}}$ has the following convergence:

\vspace{-0.35in}
$$\Delta_2^{p_1^{\circ},p_2^{\circ}}=\sum_{i=1}^{2} \pi_i~ P[{\sf d}_2^{p_1^{\circ},p_2^{\circ}}(\Xvec) \neq i\mid \Xvec \sim f(\cdot,p_i^{\circ})] \rightarrow \sum_{i=1}^{2} \pi_i~ P[{\sf d}_B (\Xvec)\neq i \mid \Xvec \sim f(\cdot, p_i^{\circ})]=\Delta_B, $$

\vspace{-0.175in}
\noindent
the Bayes risk as $\min\{n_1,n_2\} \rightarrow \infty$. 

From Theorem 1, we have ${\hat p_{1n_1}} \stackrel{P}{\rightarrow} p_1^{\circ}$ and  ${\hat p_{2n_2}} \stackrel{P}{\rightarrow} p_2^{\circ}$ as $n \to \infty$. Since ${\cal P}$ is finite, this implies
$P( {\hat p_{1n_1}}=p_1^{\circ}, {\hat p_{2n_2}} =p_2^{\circ}) \rightarrow 1$ as $n \rightarrow \infty$. The misclassification probability of the classifier ${\sf d}_2$ can be expressed as follows:
$$\sum_{p_1 \in {\cal P}} \sum _{p_2 \in {\cal P}} \Delta_2^{p_1,p_2}~ P( {\hat p_{1n_1}}=p_1, {\hat p_{1n_1}}=p_2).$$
It is now straight forward to prove that the expression above converges to $\Delta_2^{p_1^{\circ},p_2^{\circ}}(= \Delta_B, \mbox{ the Bayes risk})$ as $\min\{n_1,n_2\} \to \infty$. \hfill$\Box$

%%%%%%%%%%%%%%%%%%%%%%%%%%%%%%%%%%%%%%%%%%%%%%%%%%%%%%%%%%%%%%%%%%%%%%%%%%%%%%%%%%%%%%%%%%%%%%%%%%%%%%%%%%%%%%%%%%%%%%%%%%%%
%\noindent {\large\bf Acknowledgment}
%Write the acknowledgment here.
%\par
%\section*{Acknowledgements}

%To numerous unknown referees whose comments have improved this paper.

%%%%%%%%%%%%%%%%%%%%%%%%%%%%%%%%%%%%%%%%%%%%%%%%%%%%%%%%%%%%%%%%%%%%%%%%%%%%%%%%%%%%%%%%%%%%%%%%%%%%%%%%%%%%%%%%%%%%%%%%%%%%
%\bibliographystyle{elsarticle-num}
%\bibliography{<reference>}

%\end{document}
%\noindent{\large\bf References}
%\section*{References}
%{\small
%\newpage
%\bibliographystyle{apalike}
%\bibliography{reference}
%\begin{description}

%}

%%%%%%%%%%%%%%%%%%%%%%%%%%%%%%%%%%%%%%%%%%%%%%%%%%%%%%%%%%%%%%%%%%%%%%%%%%%%%%%%%%%%%%%%%%%%%%%%%%%%%%%%%%%%%%%%%%%%%%%%%%%%

%%%%%%%%%%%%%%%%%%%%%%%%%%%%%%%%%%%%%%%%%%%%%%%%%%%%%%%%%%%%%%%%%%%%%%%%%%%%%%%%%%%%%%%%%%%%%%%%%%%%%%%%%%%%%%%%%%%%%%%%%%%%
%%%%%%%%%%%%%%%%%%%%%%%%%%%%%%%%%%%%%%%%%%%%%%%%%%%%%%%%%%%%%%%%%%%%%%%%%%%%%%%%%%%%%%%%%%%%%%%%%%%%%%%%%%%%%%%%%%%%%%%%%%%%

\end{document}